\documentclass[a4paper,12pt]{article}
\usepackage{epsfig}
\usepackage{color}
\usepackage[numbers,sort&compress]{natbib}

\newlength{\dinwidth}
\newlength{\dinmargin}
\begin{document}
\title{  Studying of $B_{s}\to
K^{(*)-}\pi^{+},K^{(*)-}\rho^{+}$ decays within supersymmetry }
\author{ Ru-Min Wang$^1$\thanks{E-mail: ruminwang@gmail.com},
  Yuan-Guo Xu$^2$\thanks{E-mail: yuangx@iopp.ccnu.edu.cn}
  \\
{\scriptsize \it $^1$College of Physics and Electronic Engineering,
Xinyang Normal University,
 Xinyang, Henan 464000, China
}
\\
 {\scriptsize \it  $^2$Institute of Particle Physics,
 Huazhong Normal University,  Wuhan, Hubei 430079, China }
 }

 \maketitle\vspace{-1cm}
\begin{abstract}
Recent results from CDF Collaboration favor a large CP asymmetry in
$B_s\rightarrow K^-\pi^+$ decay, while the Standard Model prediction
is very small. Moreover, the measurement of its branching ratio  is
lower than the Standard Model prediction based on the QCD
factorization. We compute the gluino-mediated supersymmetry
contributions to $B_{s}\to K^{(*)-}\pi^{+}$,$K^{(*)-}\rho^{+}$
decays in the frame of the mass insertion method, and find that for
$\frac{m^2_{\tilde{g}}}{m^2_{\tilde{q}}}\leq2$,  the theoretical
predictions including the LR and RL mass insertion contributions are
compatible with the measurements of $B_s\rightarrow K^-\pi^+$ decay
and $B^0-\bar{B}^0$ mixing within $2\sigma$ ranges. Using the
constrained  LR and RL mass insertion parameter spaces, we explore
the supersymmetry mass insertion effects on the branching ratios,
the direct CP asymmetries and the polarization fractions in
$B_{s}\to K^{*-}\pi^+, K^{-}\rho^+, K^{*-}\rho^+$ decays. We find
the constrained  LR and RL insertions can provide sizable
contributions to the branching ratios of $B_{s}\to
K^{*-}\pi^{+}$,$K^{(*)-}\rho^{+}$ as well as the direct CP asymmetry
and the longitudinal polarization of $B_{s}\to K^{*-}\rho^{+}$ decay
without conflict with all related data within $2\sigma$ ranges.
  Near future experiments at Fermi Lab and CERN LHC-b can test
our predictions and shrink/reveal the mass insertion parameter
spaces.
\end{abstract}

\vspace{0cm} \noindent {\bf PACS Numbers:  12.60.Jv,
  12.15.Ji, 12.38.Bx, 13.25.Hw}

\newpage
\section{Introduction}
In the recent ten years, the successful running of $B$ factories
BABAR and Belle has provided rich experimental data for $B^{\pm}$
and $B^{0}$, which has confirmed the Kobayashi-Maskawa CP asymmetry
mechanism in the Standard Model (SM) and also shown hints for new
physics (NP).  Among the rich phenomena of $B$ decays, the decay
modes of $B$ mesons into pairs of charmless mesons  are the known
effective probes of the CP violation in the SM and  are sensitive to
potential NP scenarios beyond the SM. The two body charmless $B_s$
decays will play the similar role in
 studying the CP asymmetries (CPA), determining CKM matrix elements  and constraining/searching for
 the indirect  effects of various NP scenarios.
Recently the CDF Collaboration at Fermilab Tevatron has made the
first measurement of charmless two-body $B_{s}\rightarrow K^-\pi^+$
decay \cite{Abulencia:2006psa,Morello2008, Aaltonen2008,updated
data}
\begin{eqnarray}
&&\mathcal{B}(B_s \rightarrow K^-\pi^+)=(5.0\pm0.7\pm0.8)\times10^{-6},\nonumber\\
&&\mathcal{A}_{CP}^{dir}(B_s \rightarrow
K^-\pi^+)=0.39\pm0.15\pm0.08. \label{Eq:data}
\end{eqnarray}
The measurement is important for understanding  $B_{s}$ physics, and
also implies that many $B_{s}$ decay modes could be precisely
measured at the LHC-b.

Compared with  the  theoretical predictions for these quantities in
Refs.  \cite{Beneke:2003zv,Ali:2007ff,Williamson:2006hb}, based on
the QCD factorization (QCDF) \cite{BBNS}, the perturbative QCD
(PQCD)  \cite{PQCD}, and the soft-collinear effective theory (SCET)
\cite{SCET}, respectively,  one would find the experimental
measurement of this branching ratio agrees with the SM predictions
with SCET \cite{Williamson:2006hb},  but lower than the predictions
with QCDF and PQCD \cite{Beneke:2003zv,Ali:2007ff}. For the  CDF
measurement of $\mathcal{A}_{CP}^{dir}(B_s \rightarrow K^-\pi^+)$,
its central value favors a large CP violation in $B_s \rightarrow
K^-\pi^+$ decay (different from 0 at $2.3\sigma$), although it is
also compatible with zero. In Refs.
\cite{Gronau:2000md,Lipkin:2005pb}, a robust test of the SM or a
probe of NP is suggested by comparison of the direct CP asymmetry in
$B_s\rightarrow K^-\pi^+$ decay.

The  decays $B_s\rightarrow K^{(*)-}\pi^{+}$, $K^{(*)-}\rho^{+}$
have  been extensively studied in the literatures (for example,
Refs.
\cite{Beneke:2003zv,Beneke:2006hg,Ali:2007ff,Gronau:2000md,Lipkin:2005pb,
Williamson:2006hb,Chen:2001sx,Li:2003hea,Xu:2009,Cheng:2009mu}). The
tree-dominated decays $B_s\to K^{(*)-}\pi^{+}$, $ K^{(*)-}\rho^{+}$
are induced by $\bar{b}\to \bar{u}u\bar{d}$ transition at the quark
level, where the direct CPA are expected to be small in the SM. At
present, among many measurements in the similar modes of $B_{u,d}$
decays, several discrepancies with the SM predictions have appeared
in tree-dominated $\overline{b}\rightarrow \bar{u}u\bar{q}~(q=s,d)$
processes, for example, $B\rightarrow \pi\pi,\pi K$ puzzles
 \cite{:2008zza,Aubert:2007mj,Abe:2004us,Aubert:2005av,Aubert:2008sb}.
 Although the discrepancies are not statistically significant,
there is an unifying similarity pointing to NP (for example,
 Refs. \cite{Baek:2006ti,Yang:2005es,Buras:2004th}).
There could be also potential NP contributions in $B_s\to
K^{(*)-}\pi^{+}$, $K^{(*)-}\rho^{+}$ decays. The measurement given
in Eq. (\ref{Eq:data}) will afford an opportunity to
search/constrain NP scenarios beyond the SM.

Supersymmetry (SUSY) is an extension of the SM which emerges as one
of the most promising candidates for NP beyond the SM. In SUSY,
supersymmetric version of the SM contributes to the Flavor Change
Natural Current (FCNC) processes. The flavor-changing in these
processes is intrinsically tied to usual CKM-induced flavor-changing
of the SM (If that were the only new source of flavor physics, we
would say the model is minimally flavor violating). But general SUSY
is not minimally flavor violation. For general SUSY, a new source of
flavor violation is introduced by the squark mass matrices, which
usually can not be diagonalized on the same basis as the quark mass
matrices. This means gluinos (and other gaugios) will have
flavor-changing couplings to quarks and squarks, which implies the
FCNCs are mediated by gluinos and thus have strong interaction
strength. In order to analyze the phenomenology of non-minimally
flavor violating interactions in general SUSY framework, it is
helpful to rotate the effects so that they occur in squark
propagators rather than in couplings, and to parameterize them in
terms of dimensionless mass insertion (MI) parameters
$(\delta^{u,d}_{AB})_{ij}$ with $(A,B)=(L,R)$ and $(i,j=1,2,3)$.
In this paper, we work  in the usual MI approximation
\cite{Hall:1985dx,Gabbiani:1996hi}, and consider $B_s\rightarrow
K^{(*)-}\pi^{+}$, $K^{(*)-}\rho^{+}$ decays, in general SUSY models,
where flavor violation due to the gluino mediation can be important.
The chargino-stop and the charged Higgs-top loop contributions are
parametrically suppressed relative to the gluino contributions, and
thus are ignored following
\cite{Gabbiani:1988rb,Hagelin:1992tc,Gabrielli:1995bd,Gabbiani:1996hi}.
In our work, we also discuss the implications of  $B^0-\bar{B}^0$
mixing since the relevant MI parameters, that affect $B_s\rightarrow
K^{(*)-}\pi^{+}$, $K^{(*)-}\rho^{+}$ decays, enter also in
$B^0-\bar{B}^0$ mixing.
 We consider the LR, RL, LL and RR four kinds of the MIs.
 We find that for $\frac{m^2_{\tilde{g}}}{m^2_{\tilde{q}}}\leq2$,
  our predictions including the LR or RL MI
effects  are compatible with the measurements of $B_s\rightarrow
K^-\pi^+$ decay and $B^0-\bar{B}^0$ mixing within $2\sigma$ ranges,
and the constrained both LR and RL MIs could significantly affect
the polarization fractions of $B_s\rightarrow K^{*-}\rho^{+}$ decay.
While the constrained LL and RR insertions from $B^0-\bar{B}^0$
mixing can not explain the possible large CP asymmetry and the small
branching ratio of $B_s\rightarrow K^{-}\pi^{+}$  because of lacking
the gluino mass enhancement in the decay. Therefore, with the
ongoing $B$-physics at Tevatron, in particular with the onset of the
LHC-b experiment,
 we expect a wealth of  $B_s$ decay data and  measurements of
these observables  could restrict or reveal the parameter spaces of
 the LR and RL insertions in the near future.

The paper is arranged as follows.  In Sec. 2, the relevant formulas
for $B_s\rightarrow K^{(*)-}\pi^{+}$, $K^{(*)-}\rho^{+}$ decays and
$B^0-\bar{B}^0$ mixing are presented. We also tabulate the
theoretical inputs in this section. Sec. 3 deals with the numerical
results. Using our constrained  MI parameter spaces from $B_s\to
K^-\pi^+$ decay and  $B^0-\bar{B}^0$ mixing, we explore the MI
effects on the
 other observable quantities, which have not been measured yet in
 $B_s\to K^{(*)-}\pi^{+},K^{(*)-}\rho^{+}$ decays.
 Sec. 4 contains our summary and conclusion.

\section{The theoretical frame}

\subsection{ The decay amplitudes  for  $B_s\to K^{(*)-}\pi^{+}$, $K^{(*)-}\rho^{+}$ decays}
\label{BTOMM}
\subsubsection{The decay amplitudes in the SM}
  In the SM, the low energy effective Hamiltonian for
  the $b\to u\bar{u} d$ transition at the scale $\mu\sim m_{b}$ is given by   \cite{Buchalla:1995vs}
 \begin{eqnarray}
 \mathcal{H}^{SM}_{eff}(\Delta B=1)=\frac{G_F}{\sqrt{2}}\sum_{p=u, c}
 \lambda_p \Biggl(C^{SM}_1Q_1^p+C^{SM}_2Q_2^p
 +\sum_{i=3}^{10}C^{SM}_iQ_i+C^{SM}_{7\gamma}Q_{7\gamma}
 +C^{SM}_{8g}Q_{8g} \Biggl)+ \mbox{h.c.},
 \label{HeffSM}
 \end{eqnarray}
here  $\lambda_p=V_{pb}V_{pd}^* $ with  $p\in \{u,c\}$ are CKM
factors, the Wilson coefficients within the SM  $C^{SM}_i$ can be
found in Ref. \cite{Buchalla:1995vs}, and the relevant operators
$Q_i$ are given as
    \begin{eqnarray}
    && \!\!\!\! \!\!\!\! \!\!\!\! \!\!\!\! \!\!\!\! \!\!\!\!
    Q^{p}_{1}=({\bar{p}}_{\alpha}\gamma^\mu Lb_{\alpha})
               ({\bar{d}}_{\beta}\gamma_\mu L p_{\beta} ),
    \ \ \ \ \ \ \ \ \ \ \ \ \ \
    Q^{p}_{2}=({\bar{p}}_{\alpha}\gamma^\mu Lb_{\beta} )
               ({\bar{d}}_{\beta} \gamma_\mu Lp_{\alpha}), \nonumber\\
     &&  \!\!\!\! \!\!\!\! \!\!\!\! \!\!\!\! \!\!\!\! \!\!\!\!
    Q_{3}=({\bar{d}}_{\alpha}\gamma^\mu Lb_{\alpha})\sum\limits_{q^{\prime}}
           ({\bar{q}}^{\prime}_{\beta} \gamma_\mu Lq^{\prime}_{\beta} ),
    \ \ \ \ \ \ \ \ \ \
    Q_{4}=({\bar{d}}_{\beta} \gamma^\mu Lb_{\alpha})\sum\limits_{q^{\prime}}
           ({\bar{q}}^{\prime}_{\alpha}\gamma_\mu Lq^{\prime}_{\beta} ),   \nonumber \\
    &&  \!\!\!\! \!\!\!\! \!\!\!\! \!\!\!\! \!\!\!\! \!\!\!\!
    Q_{5}=({\bar{d}}_{\alpha}\gamma^\mu Lb_{\alpha})\sum\limits_{q^{\prime}}
           ({\bar{q}}^{\prime}_{\beta} \gamma_\mu Rq^{\prime}_{\beta} ),
    \ \ \ \ \ \ \ \ \ \
    Q_{6}=({\bar{d}}_{\beta} \gamma^\mu Lb_{\alpha})\sum\limits_{q^{\prime}}
           ({\bar{q}}^{\prime}_{\alpha}\gamma_\mu Rq^{\prime}_{\beta} ), \nonumber\\
    &&  \!\!\!\! \!\!\!\! \!\!\!\! \!\!\!\! \!\!\!\! \!\!\!\!
    Q_{7}=\frac{3}{2}({\bar{d}}_{\alpha}\gamma^\mu Lb_{\alpha})
           \sum\limits_{q^{\prime}}e_{q^{\prime}}
           ({\bar{q}}^{\prime}_{\beta} \gamma_\mu Rq^{\prime}_{\beta} ),    \ \ \ \
    Q_{8}=\frac{3}{2}({\bar{d}}_{\beta} \gamma^\mu Lb_{\alpha})
           \sum\limits_{q^{\prime}}e_{q^{\prime}}
           ({\bar{q}}^{\prime}_{\alpha}\gamma_\mu Rq^{\prime}_{\beta} ), \nonumber\\
    &&  \!\!\!\! \!\!\!\! \!\!\!\! \!\!\!\! \!\!\!\! \!\!\!\!
    Q_{9}=\frac{3}{2}({\bar{d}}_{\alpha}\gamma^\mu Lb_{\alpha})
           \sum\limits_{q^{\prime}}e_{q^{\prime}}
           ({\bar{q}}^{\prime}_{\beta}\gamma_\mu L q^{\prime}_{\beta} ),
    \ \ \ \
    Q_{10}=\frac{3}{2}({\bar{d}}_{\beta} \gamma^\mu Lb_{\alpha})
           \sum\limits_{q^{\prime}}e_{q^{\prime}}
           ({\bar{q}}^{\prime}_{\alpha}\gamma_\mu Lq^{\prime}_{\beta} ), \nonumber\\
    &&\!\!\!\! \!\!\!\! \!\!\!\! \!\!\!\! \!\!\!\! \!\!\!\!
    Q_{7{\gamma}}=\frac{e}{8{\pi}^{2}}m_{b}{\bar{d}}_{\alpha}
           {\sigma}^{{\mu}{\nu}}R
            b_{\alpha}F_{{\mu}{\nu}},
     \ \ \ \ \ \ \ \ \ \ \ \ \
    Q_{8g}=\frac{g_s}{8{\pi}^{2}}m_{b}{\bar{d}}_{\alpha}
           {\sigma}^{{\mu}{\nu}}R
            T^{a}_{{\alpha}{\beta}}b_{\beta}G^{a}_{{\mu}{\nu}},
            \label{Eq:SMoperator}
    \end{eqnarray}
where $\alpha$ and $\beta$ are color indices, and $L(R)=(1-(+)
\gamma_5)$.

 With the weak effective Hamiltonian given in Eq.
(\ref{HeffSM}),  one can  write the decay amplitudes for the
relevant two-body hadronic
 $B\to M_{1}M_{2}$ decays as
\begin{eqnarray}
  \mathcal{A}^{SM}(B\to M_1M_2)&=&\left< M_1M_2|
  {\cal H}^{SM}_{eff}(\Delta B=1)|B \right> \nonumber\\
  &=&\sum_p \sum_i \lambda_p
  C^{SM}_i(\mu)\left<M_1M_2|Q_i(\mu)|B\right>.
  \end{eqnarray}
The essential theoretical difficulty for obtaining the decay
amplitude arises  from the  evaluation of hadronic matrix elements
$\langle M_1M_2|Q_i(\mu)|B\rangle$, for which we will employ the
QCDF \cite{BBNS} throughout  this paper.
 We will use the QCDF amplitudes of these decays derived in the comprehensive papers
 \cite{Beneke:2003zv,Beneke:2006hg} as  inputs for the SM amplitudes.

\subsubsection{SUSY effects in the decays}

In SUSY extension of the SM with conserved R-parity, the potentially
most important contributions to Wilson coefficients of penguins in
the effective Hamiltonian arise from strong-interaction penguin and
box diagrams with gluino-squark loops. They can contribute to the
FCNC processes because the gluinos have flavor-changing coupling to
the quark and squark eigenstates. In general SUSY, we only consider
these potentially large gluino box and penguin contributions and
neglect a multitude of other diagrams, which are parametrically
suppressed by small electroweak gauge coupling
\cite{Gabbiani:1988rb,Hagelin:1992tc,Gabrielli:1995bd,Gabbiani:1996hi}.
The relevant Wilson coefficients of $b\to u\bar{u}d$ process  due to
the gluino box or penguin diagram involving the LL and LR insertions
are given (at the scale $\mu \sim m_W\sim m_{\tilde q}$) by
\cite{Gabbiani:1996hi,Baek:2001kc,Kane:2002sp,Ghosh:2002jpa}
\begin{eqnarray}
C_3^{SUSY}(m_{\tilde q})&=&-\frac{\alpha^2_s(m_{\tilde
q})}{2\sqrt{2}G_F\lambda_tm^2_{\tilde{q}}}
\left(-\frac{1}{9}B_1(x)-\frac{5}{9}B_2(x)-\frac{1}{18}P_1(x)-\frac{1}{2}P_2(x)\right)(\delta^d_{LL})_{13},\nonumber\\
C_4^{SUSY}(m_{\tilde q})&=&-\frac{\alpha^2_s(m_{\tilde
q})}{2\sqrt{2}G_F\lambda_tm^2_{\tilde{q}}}
\left(-\frac{7}{3}B_1(x)+\frac{1}{3}B_2(x)+\frac{1}{6}P_1(x)+\frac{3}{2}P_2(x)\right)(\delta^d_{LL})_{13},\nonumber\\
C_5^{SUSY}(m_{\tilde q})&=&-\frac{\alpha^2_s(m_{\tilde
q})}{2\sqrt{2}G_F\lambda_tm^2_{\tilde{q}}}
\left(\frac{10}{9}B_1(x)+\frac{1}{18}B_2(x)-\frac{1}{18}P_1(x)-\frac{1}{2}P_2(x)\right)(\delta^d_{LL})_{13},\nonumber\\
C_6^{SUSY}(m_{\tilde q})&=&-\frac{\alpha^2_s(m_{\tilde
q})}{2\sqrt{2}G_F\lambda_tm^2_{\tilde{q}}}
\left(-\frac{2}{3}B_1(x)+\frac{7}{6}B_2(x)+\frac{1}{6}P_1(x)+\frac{3}{2}P_2(x)\right)(\delta^d_{LL})_{13},\nonumber\\
C_{7\gamma}^{SUSY}(m_{\tilde q})&=&\frac{8\pi\alpha_s(m_{\tilde
q})}{9\sqrt{2}G_F\lambda_tm^2_{\tilde{q}}}
\left[(\delta^d_{LL})_{13}M_4(x)-(\delta^d_{LR})_{13}\left(\frac{m_{\tilde{g}}}{m_b}\right)4B_1(x)\right],\nonumber\\
C_{8g}^{SUSY}(m_{\tilde q})&=&-\frac{2\pi\alpha_s(m_{\tilde
q})}{\sqrt{2}G_F\lambda_tm^2_{\tilde{q}}}
\left[(\delta^d_{LL})_{13}\left(\frac{3}{2}M_3(x)-\frac{1}{6}M_4(x)\right)\right.\nonumber\\
&&\ \ \ \ \ \ \ \ \ \ \ \ \  \ \ \ \ \ \
\left.+(\delta^d_{LR})_{13}\left(\frac{m_{\tilde{g}}}{m_b}\right)\frac{1}{6}\left(4B_1(x)-9x^{-1}B_2(x)\right)\right],\label{Eq.SUSYWC}
\end{eqnarray}
where $x\equiv m^2_{\tilde{g}}/m^2_{\tilde{q}}$, and the loop
functions $B_i(x),P_i(x),M_i(x)$ can be found in Ref.
\cite{Baek:2001kc}. For the RR and RL insertions, we have additional
operators $\tilde{Q}_{i=3\ldots6,7\gamma,8g}$ that are obtained by
$L\leftrightarrow R$ in the SM operators  given in Eq.
(\ref{Eq:SMoperator}). The associated Wilson coefficients
$\widetilde{C}^{SUSY}_{i=3\ldots6,7\gamma,8g}$ are either dominated
by their expressions as above with the replacement $L\leftrightarrow
R$. The remaining coefficients are either dominated by their SM
$(C_{1,2})$ or are electroweak penguins $(C_{7\ldots10})$ and
therefore small.

The  SUSY Wilson coefficients at low energy $C^{SUSY}_i(\mu\sim
m_b)$ can be obtained from $C^{SUSY}_i(m_{\tilde q})$ in Eq.
(\ref{Eq.SUSYWC})  by using the renormalization group equation as
discussed in Ref. \cite{Buchalla:1995vs}
\begin{eqnarray}
C(\mu)=U_5(\mu,m_{\tilde q})C(m_{\tilde q}),
\end{eqnarray}
where $C$ is the $6\times1$ column vector of the Wilson coefficients
and $U_5(\mu, m_{\tilde q})$ is the five-flavor $6\times6$ evolution
matrix. The detailed explicitness of $U_5(\mu, m_{\tilde q})$ is
given in Ref. \cite{Buchalla:1995vs}.
The coefficients $C^{SUSY}_{7\gamma}$ and $C^{SUSY}_{7g}$ at the
$\mu\sim m_b$ scale are given by  \cite{Buras:1999da,He:2001kn}
\begin{eqnarray}
C^{SUSY}_{7\gamma}(\mu)&=&\eta^2C^{SUSY}_{7\gamma}(m_{\tilde q})+\frac{8}{3}(\eta-\eta^2)C^{SUSY}_{8g}(m_{\tilde q}),\nonumber\\
C^{SUSY}_{8g}(\mu)&=&\eta C^{SUSY}_{8g}(m_{\tilde q}),
\end{eqnarray}
where $\eta=(\frac{\alpha_s(m_{\tilde
q})}{\alpha_s(m_t)})^{\frac{2}{21}}(\frac{\alpha_s(m_t)}{\alpha_s(m_b)})^{\frac{2}{23}}$.

\subsubsection{The total decay amplitudes}

For the LL and LR insertion, the NP effective operators have the
same chirality with ones of the SM, so the total decays amplitudes
can be obtained from the SM ones in Refs.
\cite{Beneke:2003zv,Beneke:2006hg} by replacing
\begin{eqnarray}
C^{SM}_i\rightarrow C^{SM}_i+C^{SUSY}_i.
\end{eqnarray}
For RL and RR insertion, the NP effective operators have the
opposite chirality with the SM ones, and we can get the
corresponding decay amplitudes from the SM decay amplitudes by
following replacements \cite{Kagan:2004ia}
\begin{eqnarray}
C^{SM}_i\rightarrow C^{SM}_i-\widetilde{C}^{SUSY}_i,
\end{eqnarray}
for $A(B_s\rightarrow K^-\pi^+)$  and
$A_{0,\parallel}(B_s\rightarrow K^{*-}\rho^+)$, as well as
\begin{eqnarray}
C^{SM}_i\rightarrow C^{SM}_i+\widetilde{C}^{SUSY}_i,
\end{eqnarray}
for $A(B_s\rightarrow K^{*-}\pi^+),$ $A(B_s\rightarrow K^-\rho^+)$
and $A_{\perp}(B_s\rightarrow K^{*-}\rho^+)$.

Then the total branching ratios read
\begin{eqnarray}
\mathcal{B}(B_{s}\to M_1 M_2)=\frac{\tau_{B_s} |p_c |}{8\pi
m_{B_s}^2}\left|\mathcal{A}(B_s\rightarrow M_1 M_2)\right|^2,
\end{eqnarray}
where $\tau_{B_{s}}$ is the $B_{s}$ lifetime, $|p_c|$ is the center
of mass momentum in the center of mass frame of $B_s$ meson.

The direct CP asymmetry is defined as
\begin{eqnarray}
\mathcal{A}^{dir}_{CP}(B_s\rightarrow
f)=\frac{\mathcal{B}(\bar{B}_s\rightarrow
\bar{f})-\mathcal{B}(B_s\rightarrow
f)}{\mathcal{B}(\bar{B}_s\rightarrow
\bar{f})+\mathcal{B}(B_s\rightarrow f)}.
\end{eqnarray}

In the $B_s\to VV$ decay, the two vector mesons have the same
helicity, therefore three different polarization states are
possible, one longitudinal and two transverse, and we define the
corresponding amplitudes as $\mathcal{A}_{0,\pm}$.  Transverse
$(\mathcal{A}_{\parallel,\perp})$ and helicity $(\mathcal{A}_{\pm})$
amplitudes are related by
$\mathcal{A}_{\parallel,\perp}=\frac{\mathcal{A}_+\pm\mathcal{A}_-}{\sqrt{\mathcal{A}}}$.
Then we have
\begin{eqnarray}
\left|\mathcal{A}(B_s\to
VV)\right|^2&=&|\mathcal{A}_0|^2+|\mathcal{A}_+|^2+|\mathcal{A}_-|^2
=|\mathcal{A}_0|^2+|\mathcal{A}_\parallel|^2+|\mathcal{A}_\perp|^2.
\end{eqnarray}
The longitudinal(transverse) polarization fraction $f_L$($f_\perp$)
are defined by
\begin{eqnarray}
f_{L,\perp}(B_s\to
VV)&=&\frac{\Gamma_{L,\perp}}{\Gamma}=\frac{|\mathcal{A}_{0,\perp}|^2}
{|\mathcal{A}_0|^2+|\mathcal{A}_\parallel|^2+|\mathcal{A}_\perp|^2}.
\end{eqnarray}

\subsection{Effective Hamiltonian for $B^0-\bar{B}^0$ mixing}
\label{SEC.Mixing}

 The most general $B^0-\bar{B}^0$ mixing is
described by the effective Hamiltonian \cite{Becirevic:2001jj}
\begin{eqnarray}
\mathcal{H}_{eff}(\Delta
B=2)=\sum^5_{i=1}C'_iQ'_i+\sum^3_{i=1}\widetilde{C}'_i\widetilde{Q}'_i+h.c.,\label{HB2eff}
\end{eqnarray}
with
\begin{eqnarray}
Q'_1&=&(\bar{d}\gamma^\mu P_Lb)_1(\bar{d}\gamma_\mu P_Lb)_1,\nonumber\\
Q'_2&=&(\bar{d} P_Lb)_1(\bar{d} P_Lb)_1,\nonumber\\
Q'_3&=&(\bar{d} P_Lb)_8(\bar{d} P_Lb)_8,\nonumber\\
Q'_4&=&(\bar{d} P_Lb)_1(\bar{d} P_Rb)_1,\nonumber\\
Q'_5&=&(\bar{d} P_Lb)_8(\bar{d} P_Rb)_8, \label{Q}
\end{eqnarray}
where $P_{L(R)}=(1-(+)\gamma_5)/2$ and the operators
$\widetilde{Q}'_{1,2,3}$ are obtained from $Q'_{1,2,3}$ by the
exchange $L\leftrightarrow R$. The hadronic matrix elements, taking
into account for renormalization effects, are defined as
\begin{eqnarray}
\langle \bar{B}^0|
Q'_1(\mu)|B^0\rangle&=&\frac{2}{3}m^2_{B_d}f^2_{B_d}B_1(\mu),\nonumber\\
\langle \bar{B}^0|
Q'_2(\mu)|B^0\rangle&=&-\frac{5}{12}m^2_{B_d}f^2_{B_d}S_{B_d}B_2(\mu),\nonumber\\
\langle \bar{B}^0|
Q'_3(\mu)|B^0\rangle&=&\frac{1}{12}m^2_{B_d}f^2_{B_d}S_{B_d}B_3(\mu),\nonumber\\
\langle \bar{B}^0|
Q'_4(\mu)|B^0\rangle&=&\frac{1}{2}m^2_{B_d}f^2_{B_d}S_{B_d}B_4(\mu),\nonumber\\
\langle \bar{B}^0|
Q'_5(\mu)|B^0\rangle&=&\frac{1}{6}m^2_{B_d}f^2_{B_d}S_{B_d}B_5(\mu),\label{EM}
\end{eqnarray}
with
$S_{B_d}=\left(\frac{m_{B_d}}{\overline{m}_b(\overline{m}_b)+\overline{m}_d(\overline{m}_b)}\right)^2$.

The Wilson coefficients $C'_i$ receive contributions from both the
SM and the SUSY loops: $C'_i\equiv C'^{SM}_i+C'^{SUSY}_i$. In the
SM, the $t-W$ box diagram generates only contribution to the
operator $Q'_1$, and the corresponding Wilson coefficient
$C'^{SM}_1$ at the $m_b$ scale is \cite{Buchalla:1995vs}
\begin{eqnarray}
C'^{SM}_1(m_{b})=\frac{G_F^2}{4
\pi^2}m_W^2(V_{td}V^*_{tb})^2\eta_{2B}S_0(x_t)[\alpha_s(m_b)]^{-6/23}
\left[1+\frac{\alpha_s(m_b)}{4\pi}J_5\right],
\end{eqnarray}
where $x_t=m^2_t/m^2_W$ and $\eta_{2B}$ is the QCD correction.

The gluino-mediated SUSY contributions to  $B^0-\bar{B}^0$ mixing
and $2\beta$ in the MI approximation  have  been extensively studied
(for example, Refs.
\cite{Ko:2002ee,Becirevic:2001jj,Buchalla:2008jp,Altmannshofer:2009ne}).
In general SUSY models, there are new contributions to
$B^0-\bar{B}^0$ mixing from the gluino-squark box diagrams, and the
corresponding Wilson  coefficients $C'^{SUSY}_i$ (at the $m_{\tilde
q}$ scale) are given by
\cite{Gabbiani:1988rb,Hagelin:1992tc,Gabrielli:1995bd,Gabbiani:1996hi}
{\small \begin{eqnarray}
C'^{SUSY}_1(m_{\tilde q})&=&-\frac{\alpha_s^2}{216m^2_{\tilde{q}}}\left(24xf_6(x)+66\tilde{f}_6(x)\right)(\delta^d_{LL})^2_{13},\nonumber\\
C'^{SUSY}_2(m_{\tilde q})&=&-\frac{\alpha_s^2}{216m^2_{\tilde{q}}}204xf_6(x)(\delta^d_{RL})^2_{13},\nonumber\\
C'^{SUSY}_3(m_{\tilde q})&=&\frac{\alpha_s^2}{216m^2_{\tilde{q}}}36xf_6(x)(\delta^d_{RL})^2_{13},\nonumber\\
C'^{SUSY}_4(m_{\tilde
q})&=&-\frac{\alpha_s^2}{216m^2_{\tilde{q}}}\left[\left(504xf_6(x)-72\tilde{f}_6(x)\right)(\delta^d_{LL})_{13}(\delta^d_{RR})_{13}
-132\tilde{f}_6(x)(\delta^d_{LR})_{13}(\delta^d_{RL})_{13}\right],\nonumber\\
C'^{SUSY}_5(m_{\tilde
q})&=&-\frac{\alpha_s^2}{216m^2_{\tilde{q}}}\left[\left(24xf_6(x)+120\tilde{f}_6(x)\right)(\delta^d_{LL})_{13}(\delta^d_{RR})_{13}
-180\tilde{f}_6(x)(\delta^d_{LR})_{13}(\delta^d_{RL})_{13}\right].
\end{eqnarray}}
The loop functions $f_6(x),\tilde{f}_6(x)$ can be found in Ref.
\cite{Baek:2001kc}. The other Wilson coefficients
$\tilde{C}'^{SUSY}_{1,2,3}$ are obtained from $C'^{SUSY}_{1,2,3}$ by
exchange of $L\leftrightarrow R$.

The SUSY Wilson coefficients at the $m_b$ scale $C^{SUSY}_i( m_b)$
can be obtained by
\begin{eqnarray}
C_r(m_b)=\sum_i\sum_s\left(b^{(r,s)}_i+\eta'
c^{(r,s)}_i\right)\eta'^{a_i}C_s(m_{\tilde{q}}),
\end{eqnarray}
where $\eta'=\alpha_s(m_{\tilde{q}})/\alpha_s(m_t)$.  The magic
number $a_i$ $b^{(r,s)}_i$ and $c^{(r,s)}_i$ can be found in Ref.
\cite{Becirevic:2001jj}. Renormalization group evolution of
$\tilde{C}_{1,2,3}$ is done in the same way as for $C_{1,2,3}$.

In terms of the effective Hamiltonian in Eq. (\ref{HB2eff}), the
mixing amplitude $M_{12}$ read
\begin{eqnarray}
M_{12}=\frac{\langle B^0|\mathcal{H}_{eff}(\Delta
B=2)|\bar{B}^0\rangle}{2m_{B_d}}.
\end{eqnarray}
Then, the $B^0$ mass difference $\Delta M_d= 2\left|M_{12}\right|$,
its associated CP phase $2\beta= \mbox{arg}(M_{12})$, and $S_{\psi
K_S}=\mbox{sin}2\beta$. Experimental values of $\Delta M_d$ and
$2\beta$ are given by the Heavy Flavor Averaging Group
\cite{Barberio:2008fa}
\begin{eqnarray}
\Delta M_d&=& 0.507\pm0.005~ \mbox{ps}^{-1},\nonumber\\
\mbox{sin}2\beta&=& 0.68\pm0.03.\label{exp.DM}
\end{eqnarray}
Above experimental bounds will also be used in our numerical
results.

\subsection{Input Parameters}
\label{SEC.INPUT}

The input parameters are collected in Table \ref{Tab.input}. We have
several remarks on the input parameters:
\begin{itemize}
\item \underline{Wilson coefficients}: The SM Wilson coefficients $C^{SM}_i$ are obtained from the
expressions in Ref. \cite{Buchalla:1995vs}.

\item \underline{CKM matrix element}:  For the SM predictions,
we use the CKM matrix elements from the Wolfenstein parameters of
the latest analysis within the SM in Ref. \cite{UTfit}, and   for
the SUSY predictions, we take the CKM matrix elements in terms of
the Wolfenstein parameters of the NP generalized analysis results in
Ref. \cite{UTfit}.

\item  \underline{Masses of SUSY particles}:  When we study the
SUSY effects, we will consider each possible MI
$(\delta^d_{AB})_{13}$ for $AB=LL,LR,RL,RR$ only one  at a time,
neglecting the interferences between different insertions products,
but keeping their interferences  with the SM amplitude. We fix the
common squark masses $m_{\tilde{q}}=500$ GeV and consider three
values of  $x=0.25,1,4$ (i.e. $m_{\tilde{g}}=250,500,1000$ GeV) in
all case.
 \end{itemize}

\begin{table}[h]
\caption{Default values of the input parameters.} {\footnotesize
\begin{center}
\begin{tabular}{lr}\hline\hline
$m_W=80.398\pm 0.025~{\rm GeV},~m_{_{B_d}}=5.280~{\rm
GeV},~m_{_{B_s}}=5.366~{\rm
GeV},~\tau_{_{B_s}}=(1.472^{+0.024}_{-0.026})~{\rm ps},$\\
 $m_{_{K^{*\pm}}}=0.892~{\rm GeV},~m_{_{K^\pm}}=0.494~{\rm GeV},~m_{_{\pi^\pm}}=0.140{\rm GeV},~m_{\rho}=0.775~{\rm
GeV},$\\
$m_t=171.3^{+2.1}_{-1.6}~{\rm
GeV},~\overline{m}_b(\overline{m}_b)=(4.20\pm0.07)~{\rm
GeV},~\overline{m}_u(2{\rm GeV})=(0.0015\sim 0.0033)~{\rm
GeV},$\\
$\overline{m}_d(2{\rm GeV})=(0.0035\sim 0.0060)~{\rm
GeV},~\overline{m}_s(2{\rm GeV})=(0.105^{+0.025}_{-0.035})~{\rm
GeV}.$& \cite{PDG}\\\hline
The Wolfenstein parameters for the SM predictions: &\\
$A=0.810\pm0.013,~\lambda=0.2259\pm0.0016,~\bar{\rho}=0.154\pm0.022,~\bar{\eta}=0.342\pm0.014$.&\\
The Wolfenstein parameters for the SUSY predictions: & \\
$A=0.810\pm0.013,~\lambda=0.2259\pm0.0016,~\bar{\rho}=0.177\pm0.044,~\bar{\eta}=0.360\pm0.031$.&\cite{UTfit}\\
\hline
$f_{K}=0.160~{\rm GeV},~f_{K^*}=(0.217\pm0.005)~{\rm GeV},~f^{\perp}_{K^*}=(0.156\pm0.010)~{\rm GeV},$\\
$f_{\pi}=0.131~{\rm GeV},~f_{\rho}=(0.205\pm0.009)~{\rm GeV},~f^{\perp}_{\rho}=(0.147\pm0.010)~{\rm GeV},$&\\
$A^{B_{s}\to  K^*}_{0}(0)=0.360\pm0.034,~A_1^{B_{s}\rightarrow
K^{\ast}}(0)=0.233\pm 0.022,
~A_2^{B_{s}\rightarrow K^{\ast}}(0)=0.181\pm 0.025,$\nonumber\\
 $ V^{B_{s}\rightarrow K^{\ast}} (0) =0.311\pm 0.026,~F^{B_{s}\to K}_{0}(0)=0.30^{+0.04}_{-0.03}.$
& \cite{BallZwicky,Duplancic:2008tk}\\\hline
$f_{B_{s}}=(0.245\pm0.025)~{\rm GeV},~f_{B_{d}}=(0.200\pm0.020)~{\rm
GeV},~f_{B_d}\sqrt{\hat{B}_{B_d}}=0.225\pm0.025~{\rm GeV}.$ &
\cite{Lubicz:2008am}\\\hline
$\lambda_B=(0.46\pm0.11)$ GeV. &  \cite{Braun:2003wx}\\\hline
$\eta_{2B}=0.55\pm0.01.$&\cite{eta2B}\\\hline
$\alpha^\pi_1=0,~\alpha^\pi_2=0.20\pm0.15,~\alpha^\rho_1=0,~\alpha^\rho_2=0.1\pm0.2,$\\
$\alpha^K_1=0.2\pm0.2,~\alpha^K_2=0.1\pm0.3,~\alpha^{K^*}_1=0.06\pm0.06,~\alpha^{K^*}_2=0.1\pm0.2.$
&  \cite{Beneke:2003zv,Beneke:2006hg} \\\hline
$B_{1}^{(d)}(m_b)=0.87(4)\left(^{+5}_{-4}\right),~~B_{2}^{(d)}(m_b)=0.82(3)(4),~~B_{3}^{(d)}(m_b)=1.02(6)(9)$&\\
$B_{4}^{(d)}(m_b)=1.16(3)\left(^{+5}_{-7}\right),~~B_{5}^{(d)}(m_b)=1.91(4)\left(^{+22}_{-7}\right)$&\cite{Bparameter}\\\hline\hline
\end{tabular}
\end{center}}\label{Tab.input}
\end{table}

\section{Numerical results and analysis}

Now we are ready to present our numerical results and analysis.
First, we will show our estimations in the SM with the parameters
listed in Table \ref{Tab.input}
 and compare with the relevant experimental data. Then we
will study  the SUSY predictions for the branching ratios, the CP
asymmetries
 and the polarization fractions  in $B_s\to K^{(*)-}\pi^{+}$, $K^{(*)-}\rho^{+}$  decays. For CPA  of $B_s\to
K^{*-}\rho^+$, we will only study the longitudinal direct
 CP asymmetry ($\mathcal{A}^{L,dir}_{CP}$).

In the SM, the numerical results  with $1 \sigma$ error ranges for
the sensitive parameters  are presented in Table
\ref{Tab.SMpredictions}. The detailed error estimates corresponding
to the different types of theoretical uncertainties have been
already studied  in Refs. \cite{Beneke:2003zv,Beneke:2006hg}, and
our SM results of $\mathcal{B}$, $\mathcal{A}^{dir}_{CP}$ and $f_L$
are consistent with the ones in Refs.
\cite{Beneke:2003zv,Beneke:2006hg}.
\begin{table}[t]
\caption{The SM predictions with $1 \sigma$ error ranges of the
input  parameters for $\mathcal{B}$ (in units of $10^{-6}$),
$\mathcal{A}^{dir}_{CP}$ (in units of $10^{-2}$) and $f_{L}$ in
$B_s\to K^{(*)-}\pi^{+},K^{(*)-}\rho^+$ decays within QCDF. }
\begin{center}
\begin{tabular}
{l|c|c|c}\hline\hline
 Decay modes& $\mathcal{B}$&
 $\mathcal{A}^{dir}_{CP}$(
 $\mathcal{A}^{L,dir}_{CP}$)&$f_L$\\\hline
$B_s\rightarrow K^-\pi^+$&$[6.89,15.67]$&$[-8.11,-1.44]$ \\
$B_s\rightarrow K^{*-}\pi^+$&$[9.03,18.36]$&$[-0.64,2.01]$\\
$B_s\rightarrow K^-\rho^+$&$[14.79,38.39]$&$[-1.68,0.86]$\\
$B_s\rightarrow K^{*-}\rho^+$&$[11.85,69.69]$&$[-5.19,-1.02]$&$[0.87,0.97]$\\
\hline
\end{tabular}
\end{center}\label{Tab.SMpredictions}
\end{table}
For the color-allowed tree-dominated decays $B_s\to
K^-\pi^+$,$K^{*-}\pi^+$,$K^-\rho^+$,$K^{*-}\rho^+$, power
corrections have limited impact, and the main sources of theoretical
uncertainties in the branching ratios are the CKM matrix elements
and the form factors. Their $A^{dir}_{CP}$ and $A^{L,dir}_{CP}$ can
be predicted quite precisely, and found to be very small ($\sim
10^{-2}$) due to small penguin amplitudes. The uncertainty of
$f_L(B_s\to K^{*-}\rho^+)$ is mostly due to the uncertainties of
different form factors. Comparing the SM predictions in Table
\ref{Tab.SMpredictions} with the relevant experimental data in Eq.
(\ref{Eq:data}), we may see the present experimental data of  the
$B_s\to K^-\pi^+$ decay within $1\sigma$ ranges are not consistent
with the SM predictions  within $1 \sigma$ error ranges of the input
parameters. The experimental data of
$\mathcal{A}^{dir}_{CP}(B_s\rightarrow K^-\pi^+)$ are obviously
larger than the SM prediction, moreover, the measurement of its
branching ratio is lower than the SM prediction based on the QCDF.

 Now we turn to the gluino-mediated SUSY contributions to  $B_s\to K^{(*)-}\pi^{+}$,
$K^{(*)-}\rho^{+}$ decays in the framework of the MI approximation.
 These decays
are induced by $\bar{b}\to \bar{u}u\bar{d}$ transition at the quark
level, and we consider four kinds of MIs (LL, LR, RL and RR)
contributing to $B_s\to K^{(*)-}\pi^{+}$, $K^{(*)-}\rho^{+}$ decays.
In the SM, the very small direct CPA of these decays  come from the
weak phase of small penguin amplitudes. In order to have nonzero
CPA, we need at least two independent amplitudes with different weak
phases. In general SUSY models, we are considering, the weak phases
reside in the complex MI parameters $\delta$s and appear in the SUSY
Wilson coefficients in Eq. (\ref{Eq.SUSYWC}). These weak phases are
odd under a CP transformation.

First, we study constraints on the  MI parameters
$(\delta_{LL}^d)_{13}$,
 $(\delta_{LR}^d)_{13}$, $(\delta_{RR}^d)_{13}$ and
 $(\delta_{RL}^d)_{13}$ from $B_s\to K^{-}\pi^{+}$ decay and
$B^0-\bar{B}^0$ mixing. Using the formulas  in Sec. \ref{BTOMM} and
the input parameters in Sec. \ref{SEC.INPUT}, we may obtain the
two-body branching ratios and the direct CPA of $B_s\to
K^{(*)-}\pi^{+}$, $K^{(*)-}\rho^{+}$ decays within the framework of
SUSY by the QCDF. The mass difference and CP phase of
$B^0-\bar{B}^0$ mixing are gotten from the formulas in Sec.
\ref{SEC.Mixing} and the input parameters in Sec. \ref{SEC.INPUT}.
Noted that we take the relevant CKM matrix elements in terms of the
Wolfenstein parameters of the NP generalized analysis results in
\cite{UTfit}, which is different from the SM predictions.
In each of the MI scenarios to be discussed, we vary the mass
insertions over the range $|(\delta^d_{AB})_{13}|\leq 1$ to fully
map the parameter space. We then impose experimental constraints
from $B_s\to K^{-}\pi^{+}$ decay and  $B^0-\bar{B}^0$ mixing, which
are  shown in Eq. (\ref{Eq:data}) and Eq. (\ref{exp.DM}),
respectively.

 For the LL and RR MIs, $|(\delta_{LL}^d)_{13}|$ and $|(\delta_{RR}^d)_{13}|$
 are strongly constrained from  $B^0-\bar{B}^0$ mixing
\cite{Becirevic:2001jj,Buchalla:2008jp,Ghosh:2002jpa,Altmannshofer:2009ne}.
The effects of the constrained LL and RR insertions on $B_s\to
K^-\pi^+$ are almost negligible because of lacking the gluino mass
enhancement in the decay,  and they will not provide any significant
effect on $\mathcal{B}(B_s\to K^{-}\pi^{+})$ and
$\mathcal{A}^{dir}_{CP}(B_s\to K^{-}\pi^{+})$. The current data
within $1\sigma$ (or $2\sigma$) ranges of
$\mathcal{A}^{dir}_{CP}(B_s\to K^{-}\pi^{+})$ cannot be explained by
the LL and RR MIs.

The case of the  LR or RL insertion is very different from that of
either LL or RR insertion. The LR and RL MIs only generate
(chromo)magnetic operators $Q_{7\gamma,8g}$ and
$\tilde{Q}_{7\gamma,8g}$, respectively. Especially, the LR and RL
insertions are more strongly constrained, since their contributions
are enhanced by $m_{\tilde{g}}/m_b$ due to the chirality flip from
the gluino in the loop compared with the contribution including the
SM one. In these cases, even a small $(\delta_{LR}^d)_{13}$ or
$(\delta_{RL}^d)_{13}$ can have large effects in the decays.

\begin{figure}[t]
\begin{center}
\includegraphics[scale=0.5]{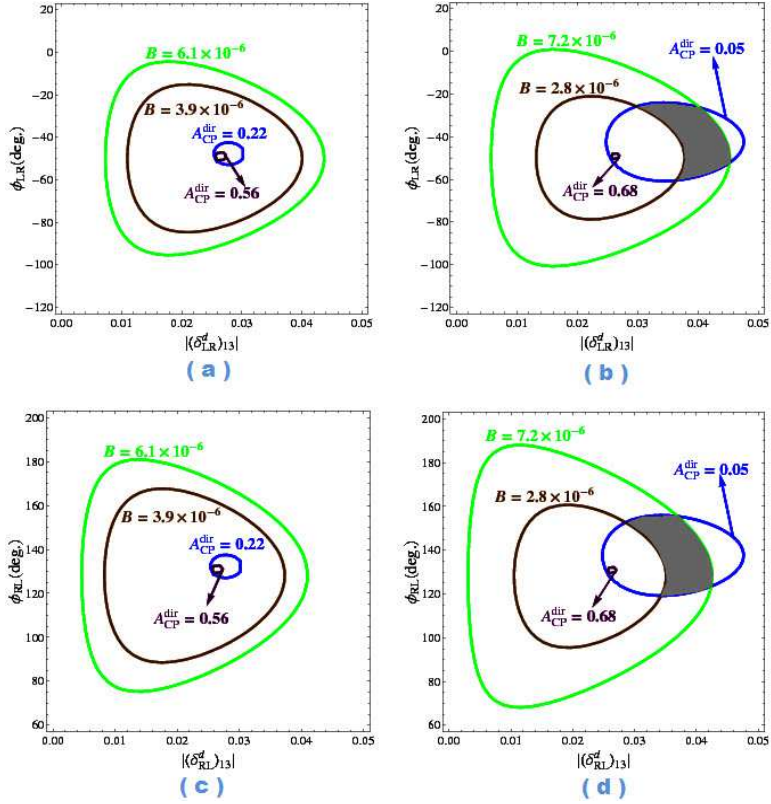}
\end{center}\vspace{-0.8cm}
\caption{ The contour plots of $\mathcal{B}(B_s\to K^{-}\pi^{+})$
and $\mathcal{A}^{dir}_{CP}(B_s\to K^{-}\pi^{+})$ in the
$|(\delta^d_{LR})_{13}|-\phi_{LR}$ and
$|(\delta^d_{RL})_{13}|-\phi_{RL}$ planes with the central values of
the input parameters and $m_{\tilde{q}}=m_{\tilde{g}}=500$ GeV.
Plots (a) and (c) show the $1\sigma$ error-bar of data given in Eq.
(\ref{Eq:data}). Plots (b) and (d) show the $2\sigma$ error-bar of
data given in Eq. (\ref{Eq:data}). $\phi_{AB}$ denotes the mixing
parameters weak phase. Please see text for details. }
 \label{fig:contour}
\end{figure}

We can not get the allowed spaces of the LR and RL insertion
parameters which may explain all these data within $1\sigma$ ranges
given in Eq.
 (\ref{Eq:data}) and Eq. (\ref{exp.DM})  at the same
 time.  In fact, there is no common allowed parameter space from $\mathcal{B}(B_s\to K^{-}\pi^{+})$ and
$\mathcal{A}^{dir}_{CP}(B_s\to K^{-}\pi^{+})$ within $1\sigma$
ranges.  Using the central values of the input parameters and
$m_{\tilde{q}}=m_{\tilde{g}}=500$ GeV,  we
 show the contour plots of $\mathcal{B}(B_s\to K^{-}\pi^{+})$ and
$\mathcal{A}^{dir}_{CP}(B_s\to K^{-}\pi^{+})$ in the
$|(\delta^d_{LR})_{13}|-\phi_{LR}$ and
$|(\delta^d_{RL})_{13}|-\phi_{RL}$ planes in Fig. \ref{fig:contour}.
We discuss Fig. \ref{fig:contour} (a) in detail. Fig.
\ref{fig:contour} (a) show the contour plot of $1\sigma$ error bars
of $\mathcal{B}(B_s\to K^{-}\pi^{+})$ and
$\mathcal{A}^{dir}_{CP}(B_s\to K^{-}\pi^{+})$  in
$|(\delta^d_{LR})_{13}|-\phi_{LR}$ plane. The space between the
contour lines of  $\mathcal{B}(B_s\to
K^{-}\pi^{+})=6.1\times10^{-6}$ and  $\mathcal{B}(B_s\to
K^{-}\pi^{+})=3.9\times10^{-6}$ is the allowed space for
$\mathcal{B}(B_s\to K^{-}\pi^{+})$ within $1\sigma$ ranges.  The
space between the contour lines of  $\mathcal{A}^{dir}_{CP}(B_s\to
K^{-}\pi^{+})=0.56$ and $\mathcal{A}^{dir}_{CP}(B_s\to
K^{-}\pi^{+})=0.22$ is the allowed space for
$\mathcal{A}^{dir}_{CP}(B_s\to K^{-}\pi^{+})$ within $1\sigma$
ranges.  From  Fig. \ref{fig:contour} (a), we can see that there is
no intersection between the allowed space from $\mathcal{B}(B_s\to
K^{-}\pi^{+})$ and the allowed space  from
$\mathcal{A}^{dir}_{CP}(B_s\to K^{-}\pi^{+})$, and  the experimental
data of  $\mathcal{A}^{dir}_{CP}(B_s\to K^{-}\pi^{+})$ within
$1\sigma$ give quite strong constraints  on $(\delta_{LR}^d)_{13}$.
The similar results for the RL insertion are shown in Fig.
\ref{fig:contour} (c) in $|(\delta^d_{RL})_{13}|-\phi_{RL}$ plane,
and there is also no intersection between the allowed space from
$\mathcal{B}(B_s\to K^{-}\pi^{+})$ within $1\sigma$ ranges and the
allowed space  from $\mathcal{A}^{dir}_{CP}(B_s\to K^{-}\pi^{+})$
within $1\sigma$ ranges.

\begin{figure}[htb]
\begin{center}
\includegraphics[scale=1.4]{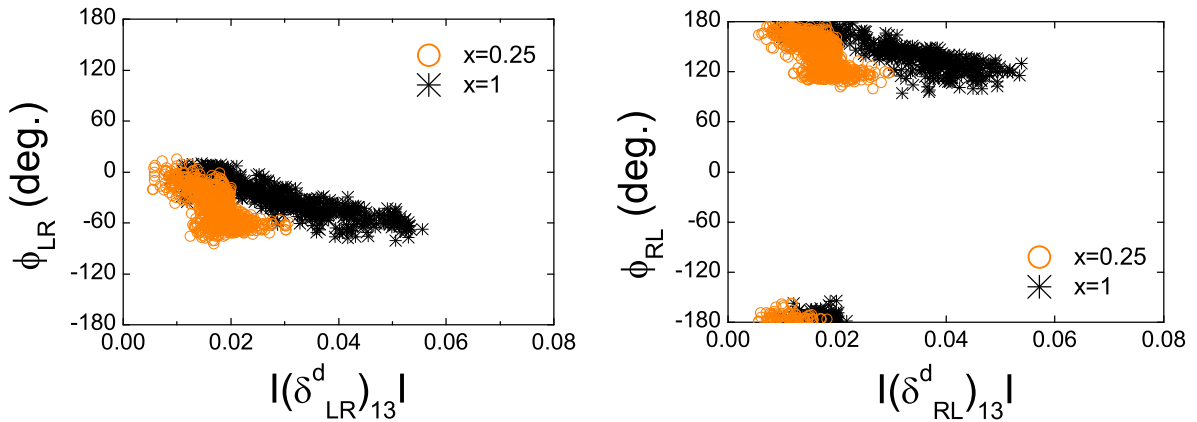}
\end{center}\vspace{-0.8cm}
\caption{ The allowed parameter spaces of the LR and RL MI
parameters constrained from $B_s\to K^{-}\pi^{+}$ decay and
$B^0-\bar{B}^0$ mixing at 95\% C.L. for the squark mass
$m_{\tilde{q}}=500$ GeV
 and for the different values of $x=0.25$ and $x=1$.}
 \label{fig:bounds}
\end{figure}

Then we expand
 the experimental bounds within $2\sigma$
 ranges to search for the allowed spaces of the LR and RL MI
 parameters. As shown in the dark gray ranges of Fig. \ref{fig:contour} (b) and
 (d), there are the allowed spaces for the LR and RL insertion
 parameters from the data of $B_s\to K^{-}\pi^{+}$ decay
within $2\sigma$ ranges. Noted that the allowed parameter spaces
shown in the dark gray ranges of Fig. \ref{fig:contour} (b) and
 (d) are obtained by using the central values of the input
 parameters. The allowed spaces will be enlarged if we consider the
 theoretical uncertainties of the input parameters.
We use the experimental bounds of $B_s\to K^{-}\pi^{+}$ decay and
$B^0-\bar{B}^0$ mixing within $2\sigma$ ranges, and take the input
parameters within $2\sigma$ ranges to obtain the allowed spaces of
the LR and RL insertion parameters.
The constrained spaces of $(\sigma_{LR}^d)_{13}$ and
$(\sigma_{RL}^d)_{13}$  for $m_{\tilde{q}}=500$ GeV and different
$x$ are demonstrated in Fig. \ref{fig:bounds}, and the corresponding
numerical ranges are summarized in Table \ref{Tab.bounds}.

\begin{table}[htb]
\caption{Bounds on the LR and RL MI parameters from the measurements
of  $B_s\to K^{-}\pi^{+}$
 decay and  $B^0-\bar{B}^0$ mixing  at 95\% C.L.. }
\begin{center}{\small
\begin{tabular}
{c|cc|cc}\hline\hline $x$&
$|(\delta_{LR}^d)_{13}|$&$\phi_{LR}$(deg.)
&$|(\delta_{RL}^d)_{13}|$& $\phi_{RL}$(deg.)\\\hline
$0.25$&$[0.005,0.031]$&$[-85,15]$&$[0.005,0.030]$&$[100,180]\cup[-180,-158]$\\
$1$&$[0.008,0.056]$&$[-81,10]$&$[0.008,0.054]$&$[94,180]\cup[-180,-154]$\\\hline
\end{tabular}}
\end{center}\label{Tab.bounds}
\end{table}

In Fig. \ref{fig:bounds}, we can see the allowed moduli of the LR
and RL MI
 parameters are very sensitive to
the values of $x$, nevertheless the allowed phase ranges of the LR
and RL MI parameters are not changed obviously for different $x$.
We find that both moduli and phases of $(\delta_{LR,RL}^d)_{13}$ are
strongly constrained by  the data of the $B_s \to K^-\pi^+$ decay
within $2\sigma$
 ranges,  since the SM
 prediction of $\mathcal{A}^{dir}_{CP}(B_s \to
 K^-\pi^+)$ with the QCDF is not  consistent with the experimental data within $2\sigma$
 ranges.  The lower limits of $|(\delta_{LR,RL}^d)_{13}|$ come from the experimental lower limit of
$\mathcal{A}^{dir}_{CP}(B_s \to
 K^-\pi^+)$ within $2\sigma$ ranges. In the case of $x=1$, $\phi_{LR,RL}$ are also
 constrained by $\Delta M_d$ within $2\sigma$ ranges, while
sin$2\beta$ within $2\sigma$ ranges doesn't provide any further
constraint.  It's different in the case of $x=0.25$, $\phi_{LR,RL}$
are also
 constrained by sin$2\beta$ within $2\sigma$ ranges, but $\Delta M_d$ within $2\sigma$ ranges doesn't provide any further
constraint.

It is worth to note that we also study the constrained spaces of
$(\delta_{LR}^d)_{13}$ and $(\delta_{RL}^d)_{13}$  in the case of
$x=4$, and we find there is no intersection of the constrained
spaces between from $B^0-B^0$ mixing within $2\sigma$
 ranges and  from $B_s \to K^-\pi^+$ decay within $2\sigma$
 ranges.
From $B^0-B^0$ mixing within $2\sigma$ ranges, we get
$|(\delta_{LR,RL}^d)_{13}|\leq 0.07$, in which $\phi_{LR,RL}
\in[-180^\circ,180^\circ]$ if $|(\delta_{LR,RL}^d)_{13}|\leq 0.03$,
and  $\phi_{LR,RL}
\in[-110^\circ,-45^\circ]\cup[70^\circ,135^\circ]$ if
$|(\delta_{LR,RL}^d)_{13}|\in[0.03,0.07]$.  We obtain
$|(\delta_{LR,RL}^d)_{13}|\in[0.03,0.19]$ from $B_s \to K^-\pi^+$
decay within $2\sigma$ ranges, in which  $\phi_{LR}
\in[-90^\circ,5^\circ]$ ($\phi_{RL}
\in[-180^\circ,-175^\circ]\cup[90^\circ,180^\circ]$) if
$|(\delta_{LR,RL}^d)_{13}|\in[0.07,0.19]$, and $\phi_{LR}
\in[-40^\circ,15^\circ]$ ($\phi_{RL}
\in[-180^\circ,-165^\circ]\cup[140^\circ,180^\circ]$) if
$|(\delta_{LR,RL}^d)_{13}|\in[0.03,0.07]$. So there is no common
allowed phases between the bounds from $B^0-B^0$ mixing
 and the bounds from $B_s \to K^-\pi^+$ decay in the case of
 $x=4$.  If $x=2$, there still are  common allowed spaces for the LR and RL MIs. For the LR MI, two
 very narrow allowed spaces are near $(|(\delta_{LR}^d)_{13}|\approx0.02,
 \phi_{LR}\approx4^\circ)$ and $(|(\delta_{LR}^d)_{13}|\approx0.04,
 \phi_{LR}\approx-64^\circ)$. For the RL MI, two
 very allowed spaces are near $(|(\delta_{RL}^d)_{13}|\approx0.02,
 \phi_{RL}\approx-176^\circ)$ and $(|(\delta_{RL}^d)_{13}|\approx0.04,
 \phi_{RL}\approx116^\circ)$.

The relevant upper bounds have been  obtained  in Refs.
\cite{Becirevic:2001jj,Buchalla:2008jp,Altmannshofer:2009ne}.  In
Ref. \cite{Becirevic:2001jj}, $|(\delta_{LR,RL}^d)_{13}|\leq
0.07,~0.08,~0.11$ for $x=0.25,~1,~4$, respectively, which are
constrained  by  imposing the experimental bounds from $\Delta M_d$
and $S_{\psi K_S}$ within $1\sigma$ ranges, setting
$m_{\tilde{q}}=500$ GeV  and scanning over the CKM phase
$\gamma\in[0,2\pi]$. In Ref. \cite{Buchalla:2008jp},
$|(\delta_{LR,RL}^d)_{13}| \leq 0.015$ for
$m_{\tilde{g}}=m_{\tilde{q}}=500$ GeV from $\triangle M_d$ and
$2\beta$ within $2\sigma$ ranges. In Ref.
\cite{Altmannshofer:2009ne}, $|(\delta_{RL,RL}^d)_{13}|\leq0.03$ for
$m_{\tilde{g}},m_{\tilde{q}}\leq600$ GeV from $\triangle M_d$ and
$2\beta$.
In the case of $x=0.25$ and $x=1$, comparing with the exist bounds
in \cite{Becirevic:2001jj,Buchalla:2008jp,Altmannshofer:2009ne}, our
upper limits of $|(\delta_{LR,RL}^d)_{13}|$ are at the same order of
previous ones, while the lower limits of $|(\delta_{LR,RL}^d)_{13}|$
are also given  from $B_s \to K^-\pi^+$ decay within $2\sigma$
ranges. In addition, $\phi_{LR,RL}$ are strongly constrained from
$B_s \to K^-\pi^+$ decay. The allowed space for $x=4$ case are ruled
out by both $B_s \to K^-\pi^+$ decay and $B^0-B^0$ mixing together.

Next, we will explore  the SUSY effects on the other quantities,
which have not been measured yet in $B_s\to
K^{*-}\pi^{+}$,$K^{(*)-}\rho^{+}$ decays, by using the constrained
parameter spaces of the LR and LR insertions as shown in Fig.
\ref{fig:bounds}.  With the expressions for $\mathcal{B}$,
$\mathcal{A}^{dir}_{CP}$ and $f_L$, we perform a scan through the
input parameters within $2\sigma$ ranges  and the new constrained
SUSY MI parameter
 spaces, and then the allowed ranges for $\mathcal{B}$,
$\mathcal{A}^{dir}_{CP}$ and $f_L$ are obtained with different SUSY
mixing insertion parameter, which satisfy relevant experimental
constraints of $B_s\to K^{-}\pi^{+}$ decay given in Eq.
(\ref{Eq:data}) and $B^0-\bar{B}^0$ mixing given in Eq.
(\ref{exp.DM}). The numerical results for $B_s \rightarrow
K^{*-}\pi^+, K^{(*)-}\rho^+$ with different $x$ value are summarized
in Table \ref{Tab.SUSYpredictions}.
\begin{table}[b]
\caption{The theoretical predictions for $\mathcal{B}$ (in units of
$10^{-6}$), $\mathcal{A}^{dir}_{CP}$ (in units of $10^{-2}$) and
$f_{L}$ in four $B_s\to K^{(*)-}\pi^+$, $K^{(*)-}\rho^{+}$ decays
based on general SUSY models with different MI and different $x$.
The corresponding SM predictions with $2\sigma$ error ranges of the
input  parameters are also listed for comparison.  }
\begin{center}{\footnotesize
\begin{tabular}
{l|c|c|c|c|c}\hline\hline
 & &
 SUSY values
 & SUSY values&SUSY values& SUSY values \\
 Observables &SM predictions &with $(\delta_{LR}^d)_{13}$&with  $(\delta_{LR}^d)_{13}$&with $(\delta_{RL}^d)_{13}$&with $(\delta_{RL}^d)_{13}$\\
 && for $x=1$&for $x=0.25$&for $x=1$&for $x=0.25$\\\hline
$\mathcal{B}(B_s\rightarrow
K^{-}\pi^+)$&$[3.32,24.48]$&$[2.80,7.20]$&$[2.80,7.20]$&$[2.80,7.20]$&$[2.80,7.20]$\\\hline
$\mathcal{B}(B_s\rightarrow
K^{*-}\pi^+)$&$[5.91,25.80]$&$[0.25,30.67]$&$[0.13,30.94]$&$[7.10,93.54]$&$[7.53,93.32]$\\\hline
$\mathcal{B}(B_s\rightarrow
K^{-}\rho^+)$&$[7.36,57.97]$&$[0.01,33.31]$&$[0.01,31.51]$&$[15.96,275.11]$&$[17.51,273.01]$\\\hline
$\mathcal{B}(B_s\rightarrow K^{*-}\rho^+)$&$[2.47,127.56]$&$[1.36,64.77]$&$[1.21,77.47]$&$[1.21,72.76]$&$[1.08,81.71]$\\
\hline
$\mathcal{A}_{CP}(B_s\rightarrow
K^{-}\pi^+)$&$[-14.46,0.97]$&$[5.00,25.07]$&$[5.00,22.02]$&$[5.00,22.73]$&$[5.00,19.10]$\\\hline
$\mathcal{A}_{CP}(B_s\rightarrow
K^{*-}\pi^+)$&$[-4.62,6.12]$&$[-12.31,4.88]$&$[-20.27,13.30]$&$[-5.15,4.84]$&$[-4.40,5.21]$\\\hline
$\mathcal{A}_{CP}(B_s\rightarrow
K^{-}\rho^+)$&$[-3.63,2.03]$&$[-36.59,13.70]$&$[-35.67,34.38]$&$[-1.85,1.48]$&$[-1.76,1.26]$\\\hline
$\mathcal{A}^{L,dir}_{CP}(B_s\rightarrow
K^{*-}\rho^+)$&$[-13.32,1.62]$&$[-80.64,36.23]$&$[-94.55,63.70]$&$[-54.42,56.75]$&$[-75.29,65.98]$\\\hline
$f_{L}(B_s\rightarrow
K^{*-}\rho^+)$&$[0.51,0.98]$&$[0.02,0.97]$&$[0.01,0.98]$&$[0.03,0.98]$&$[0.01,0.98]$\\\hline
\end{tabular}}
\end{center}\label{Tab.SUSYpredictions}
\end{table}
The corresponding SM predictions with $2\sigma$ error ranges of the
input  parameters are also listed for comparison in the second
column of the  Table \ref{Tab.SUSYpredictions}. We can see the data
of $\mathcal{B}(B_s\to K^{-}\pi^{+})$ is consistent with the SM
prediction of $\mathcal{B}(B_s\to K^{-}\pi^{+})$ at $95\%$ C.L.,
nonetheless very close  the lower limit of its SM prediction. The
data of $\mathcal{A}^{dir}_{CP}(B_s\to K^{-}\pi^{+})$ is not
consistent with its SM prediction  at $95\%$ C.L.. From the last
four columns of the Table \ref{Tab.SUSYpredictions}, we can see the
results are similar for different value of $x$. In the SUSY
predictions of the branching ratios,  there are many sources of
uncertainties, mainly arising from different form factors, CKM
matrix elements, the annihilation contribution, other hadronic
parameters of the QCDF, and the constrained MI parameters. The
uncertainties of these direct CPA mostly come from the constrained
MI parameters. The uncertainty of $f_{L}(B_s\rightarrow
K^{*-}\rho^+)$ due to different form factors and the constrained MI
parameters.

Comparing the  SUSY predictions to the SM predictions given in Table
\ref{Tab.SUSYpredictions}, we give some remarks on the numerical
results:
\begin{itemize}

\item The LR and RL MIs have significant effects on
$\mathcal{B}(B_s\rightarrow K^-\pi^+)$, and the relevant parameters
have been
 limited by both upper and lower limits of $\mathcal{B}(B_s\rightarrow
K^-\pi^+)$ within $2\sigma$ ranges. The LR and RL MIs also have
great effects on $\mathcal{A}^{dir}_{CP}(B_s\rightarrow$ $
K^-\pi^+)$, which could be increased from the SM prediction range
$[-0.14,0.01]$  to about $[0.05,0.2]$, however, the range
$[0.05,0.2]$ is still far from the central value of its measurement,
and is near to the lower limit of the measurement within $2\sigma$
ranges. The LR and RL MI parameters just have been limited by the
lower limit of $\mathcal{A}^{dir}_{CP}(B_s\rightarrow K^-\pi^+)$
within $2\sigma$ ranges.

\item The constrained LR insertion still has significant effects on
$\mathcal{B}(B_s\to K^{*-}\pi^+,K^{(*)-}\rho^{+})$. The allowed
lower limits of $\mathcal{B}(B_s\to K^{*-}\pi^+,K^{-}\rho^{+})$
could be reduced one or two order(s) from their SM predictions, and
the allowed upper limit of $\mathcal{B}(B_s\to K^{*-}\rho^{+})$ has
been suppressed a lot from its SM prediction. In the SM, the direct
CPA are very small in these decays. It is interesting to find the
contributions of the constrained LR insertion have great effects on
$\mathcal{A}^{dir}_{CP}(B_s\to K^{*-}\pi^+,K^{(*)-}\rho^{+})$. The
allowed ranges of $\mathcal{A}^{dir}_{CP}(B_s\to
K^{*-}\pi^+,K^{(*)-}\rho^{+})$ could be extremely enlarged from ones
of their tiny SM predictions.  We find the constrained LR insertion
contributions  have a great impact on the longitudinal polarization
fraction $f_{L}(B_s\rightarrow K^{*-}\rho^+)$, which could be
reduced to about zero by the constrained LR insertion.

\item For the RL MI case, the effects of the constrained RL
insertion could exceedingly increase the allowed upper limits of
$\mathcal{B}(B_s\to K^{*-}\pi^+,K^{-}\rho^{+})$ and decrease the
upper limit  of $\mathcal{B}(B_s\to K^{*-}\rho^{+})$. The RL
insertion has small effects on $\mathcal{A}^{dir}_{CP}(B_s\to
K^{*-}\pi^+,K^{-}\rho^{+})$, however, this insertion can greatly
affect $\mathcal{A}^{L,dir}_{CP}(B_s\to K^{*-}\rho^{+})$.
$f_L(B_s\to K^{*-}\rho^{+})$ also could be reduced to about zero by
the constrained RL insertion.
\end{itemize}

Noted that the LR and RL MIs only generate dipole operators
$Q_{7\gamma,8g}$ and $\tilde{Q}_{7\gamma,8g}$, respectively, and
$Q_{7\gamma,8g}$,$\tilde{Q}_{7\gamma,8g}$ do not contribute to the
transverse penguin amplitudes at $\mathcal{O}(\alpha_s)$ due to
angular momentum conservation  in $B_s\rightarrow K^{*-}\rho^+$
decay \cite{Kagan:2004uw}. In other words, the LR and RL MIs only
contribute to the longitudinal penguin amplitude at
$\mathcal{O}(\alpha_s)$. Because the LR and RL contributions are
enhanced by $m_{\tilde{g}}/m_b$, even a small $(\delta_{LR}^d)_{13}$
or $(\delta_{RL}^d)_{13}$ can have large effects on the longitudinal
penguin amplitude, and then can significantly affect the
polarization fractions of $B_s\rightarrow K^{*-}\rho^+$ decay.  For
the similar reason, the LR and RL MIs have been proposed as a
possible resolution to the polarization puzzle in $B\to \phi K^*$
decays \cite{Huang:2005qb,Giri:2004zj}.

 For each
LR and RL insertions, we can present the distributions and
correlations of $\mathcal{B}$, $\mathcal{A}^{dir}_{CP}$,
 $f_{L}$ within the modulus or weak phase of the
constrained MI parameter space in Fig. \ref{fig:bounds} by
two-dimensional scatter plots.
\begin{figure}[ht]
\begin{center}
\includegraphics[scale=0.5]{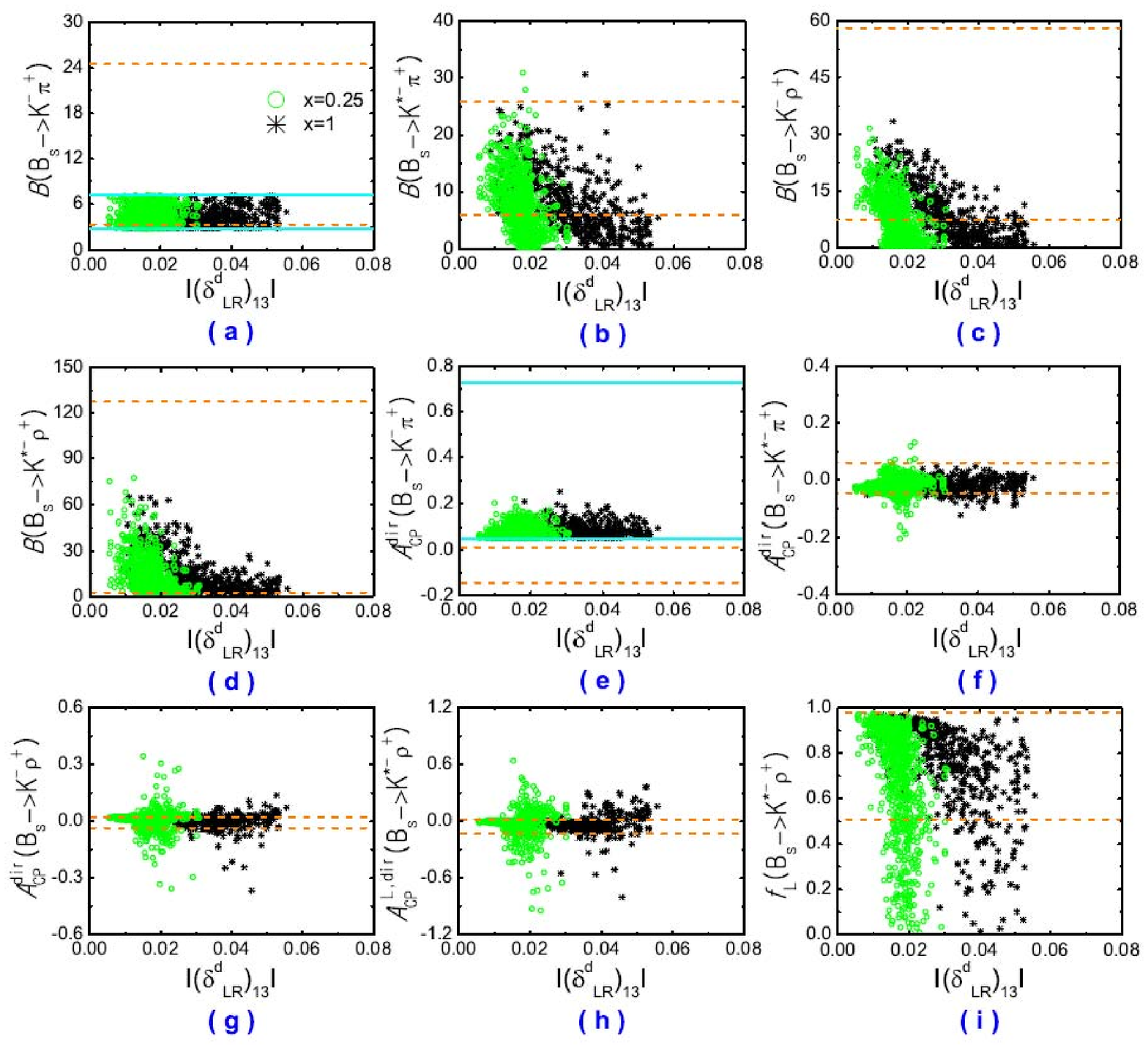}
\end{center}\vspace{-0.8cm}
\caption{\small The effects of $\left|(\delta_{LR}^d)_{13}\right|$
in $B_s\to K^{(*)-}\pi^+, K^{(*)-}\rho^{+}$ decays.
$\mathcal{A}^{dir}_{CP}$ and $\mathcal{B}$
 are in units of $10^{-2}$ and $10^{-6}$, respectively. The orange horizontal dash lines denote the limits of SM predictions,
 and the cyan horizontal solid lines represent the $2\sigma$  error bar of the measurements.
 (The same in Figs. \ref{fig:PLR}-\ref{fig:PRL}).}
 \label{fig:MLR}
\begin{center}
\includegraphics[scale=0.5]{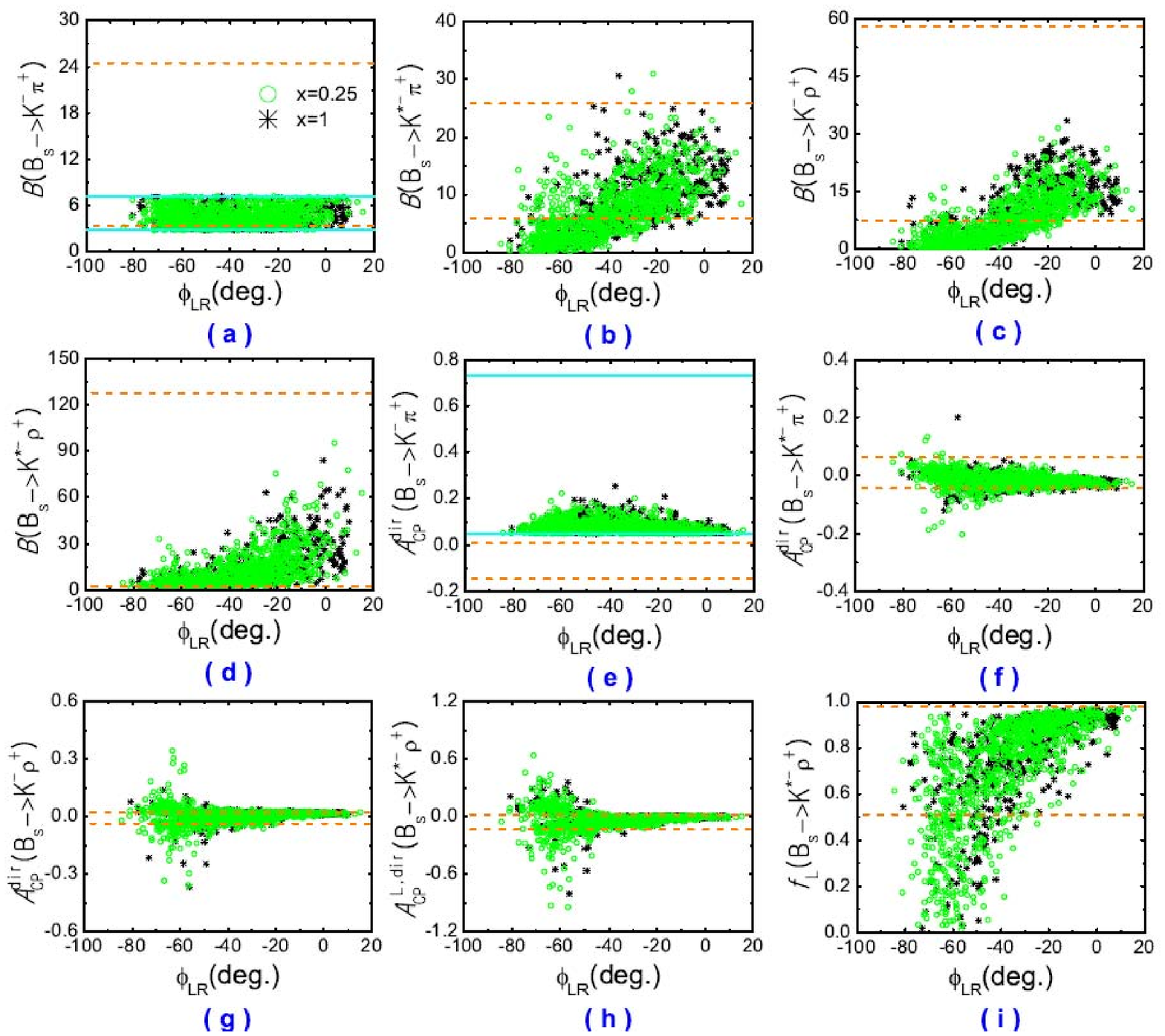}
\end{center}\vspace{-0.8cm}
\caption{\small The effects of $\phi_{LR}$ in $B_s\to K^{(*)-}\pi^+,
K^{(*)-}\rho^{+}$ decays.  }
 \label{fig:PLR}
\end{figure}
 The LR MI effects on all observables of $B_s\to K^{(*)-}\pi^+,
K^{(*)-}\rho^{+}$ are displayed in Fig. \ref{fig:MLR} and  Fig.
\ref{fig:PLR}. Fig. \ref{fig:MLR} shows the sensitivities of all
observables to $|(\delta_{LR}^d)_{13}|$ for different value of $x$,
and we see that all observables expect $\mathcal{B}(B_s\to
K^{-}\pi^+)$ are a little  sensitive to both
$|(\delta_{LR}^d)_{13}|$ and  the values of $x$. Fig. \ref{fig:PLR}
displays the sensitivities of the observables to $\phi_{LR}$ for
different $x$, and we can see that the weak phase $\phi_{LR}$ for
different $x$ value has similar allowed ranges, and has similar
effects on every observable. In addition, for comparing
conveniently, we show the SM bounds of these observables by orange
horizontal dash lines and the limits of the measurements
 of $B_s\to K^-\pi^+$ within $2\sigma$  error-bar  by  the cyan horizontal solid lines.
From Fig. \ref{fig:MLR}(a-d) and  Fig. \ref{fig:PLR}(a-d), we see
that  $\mathcal{B}(B_s\to K^{-}\pi^+)$ is strongly constrained from
its experimental data, $\mathcal{B}(B_s\to
K^{*-}\pi^+,K^{(*)-}\rho^{+})$ are very sensitive to both
$|(\delta_{LR}^d)_{13}|$ and $\phi_{LR}$, and they are decreasing
with $|(\delta_{LR}^d)_{13}|$ but increasing with $\phi_{LR}$. As
shown in  Fig. \ref{fig:MLR}(e) and  Fig. \ref{fig:PLR}(e), the LR
insertion has positive effects on $\mathcal{A}^{dir}_{CP}(B_s\to
K^{-}\pi^+)$, and there is no any point in the SM area since
$\mathcal{A}^{dir}_{CP}(B_s\rightarrow K^{-}\pi^+)$ is strongly
constrained by the corresponding experimental data, which are not
consistent with the SM predictions at 95\% C.L.. Fig. \ref{fig:MLR}
(f-h) and Fig. \ref{fig:PLR} (f-h) display that
 $\mathcal{A}^{dir}_{CP}(B_s\to K^{*-}\pi^+,
K^{(*)-}\rho^+)$ are  sensitive to $|(\delta_{LR}^d)_{13}|$, and
could have very large allowed ranges when $\phi_{LR}
\in[-80^\circ,-40^\circ]$. As for the LR insertion effects on
$f_L(B_s\to K^{*-}\rho^+)$, we show it in Fig. \ref{fig:MLR} (i) and
Fig. \ref{fig:PLR} (i), and we can see $f_L(B_s\to K^{*-}\rho^+)$
could be hugely affected by the LR MI. $f_L(B_s\to K^{*-}\rho^+)$
has some sensitivities  to both $|(\delta_{LR}^d)_{13}|$ and
$\phi_{LR}$, and it has smaller allowed range with $\phi_{LR}$. So
the future measurement of $f_L(B_s\to K^{*-}\rho^+)$ could give
obvious constraint on $\phi_{LR}$.

\begin{figure}[b]
\begin{center}
\includegraphics[scale=0.55]{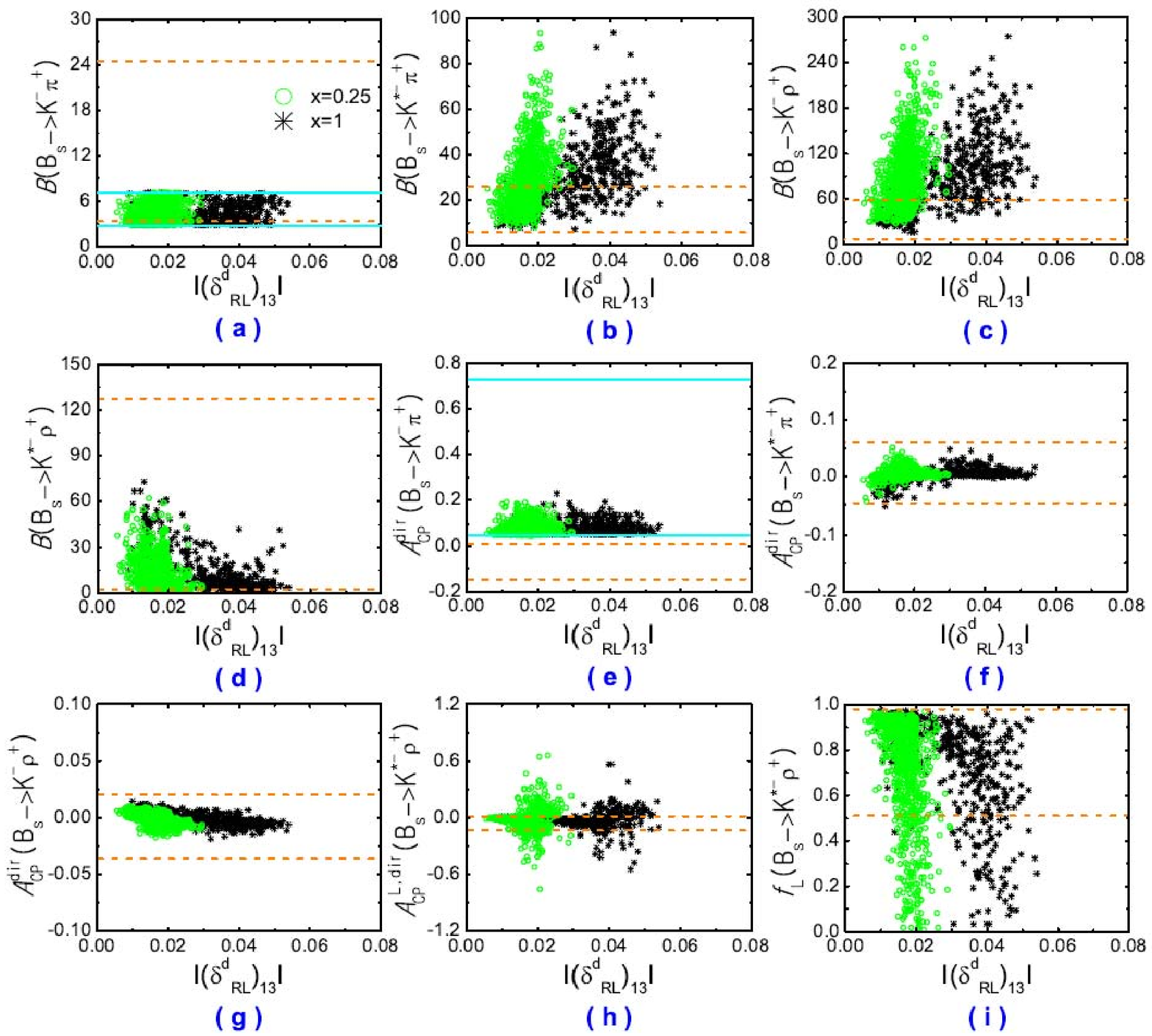}
\end{center}\vspace{-0.8cm}
 \caption{\small The effects of $\left|(\delta_{RL}^d)_{13}\right|$ in $B_s\to
K^{(*)-}\pi^+, K^{(*)-}\rho^{+}$ decays.   }
 \label{fig:MRL}
\begin{center}
\includegraphics[scale=0.55]{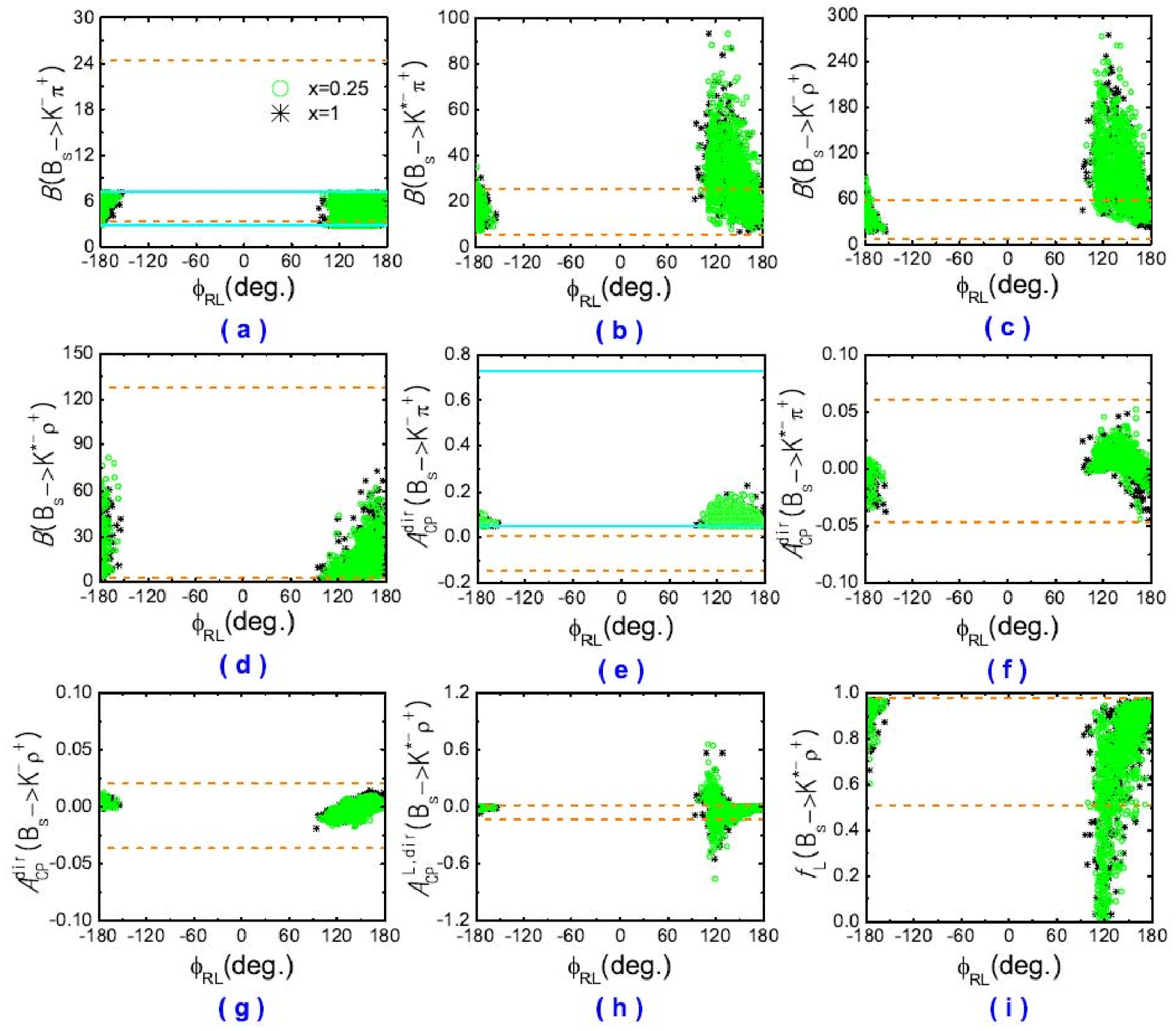}
\end{center}\vspace{-0.8cm}
 \caption{\small The effects of $\phi_{RL}$ in $B_s\to
K^{(*)-}\pi^+, K^{(*)-}\rho^{+}$ decays. }
 \label{fig:PRL}
\end{figure}
Next,  we discuss the RL MI effects on all observables in $B_s\to
K^{(*)-}\pi^+, K^{(*)-}\rho^{+}$  decays.
 Fig. \ref{fig:MRL}
and  Fig. \ref{fig:PRL} show the observables as functions of
$|(\delta_{RL}^d)_{13}|$  and $\phi_{RL}$, respectively. Fig.
\ref{fig:MRL}(a,d) and Fig. \ref{fig:PRL}(a,d) show the constrained
RL MI has negative effects on $\mathcal{B}(B_s\to
K^{-}\pi^+,K^{*-}\rho^{+})$, which is same as the LR MI effects on
them. Fig. \ref{fig:MRL}(b,c) and Fig. \ref{fig:PRL}(b,c) show us
the constrained RL MI has very large positive effects on
$\mathcal{B}(B_s\to K^{*-}\pi^+,K^{-}\rho^{+})$, which is different
from the LR MI. $\mathcal{B}(B_s\to K^{*-}\pi^+,K^{(*)-}\rho^{+})$
 have some sensitivities to $|(\delta_{RL}^d)_{13}|$  and $\phi_{RL}$
as displayed in Fig. \ref{fig:MRL}(b-d) and  Fig. \ref{fig:PRL}
(b-d). As shown in Fig. \ref{fig:MRL}(f,g) and  Fig. \ref{fig:PRL}
(f,g), the RL MI has small effects on $\mathcal{A}^{dir}_{CP}(B_s\to
K^{*-}\pi^+, K^{-}\rho^+)$, and the RL insertion contributions can
not be distinguished from  the SM prediction.
 Fig. \ref{fig:MRL} (h) and Fig.
\ref{fig:PRL} (h) show the RL MI effects on
$\mathcal{A}^{dir}_{CP}(B_s\to K^{*-}\rho^+)$ are similar to the LR
insertion effects on it. $\mathcal{A}^{dir}_{CP}(B_s\to
K^{*-}\rho^+)$ could has very large allowed ranges when $\phi_{RL}
\in[100^\circ,140^\circ]$.  As shown in Fig. \ref{fig:PRL} (i),
$f_{L}(B_s\rightarrow K^{*-}\rho^+)$ could be strongly suppressed by
the RL insertion, too.

\begin{figure}[ht]
\begin{center}
\includegraphics[scale=0.8]{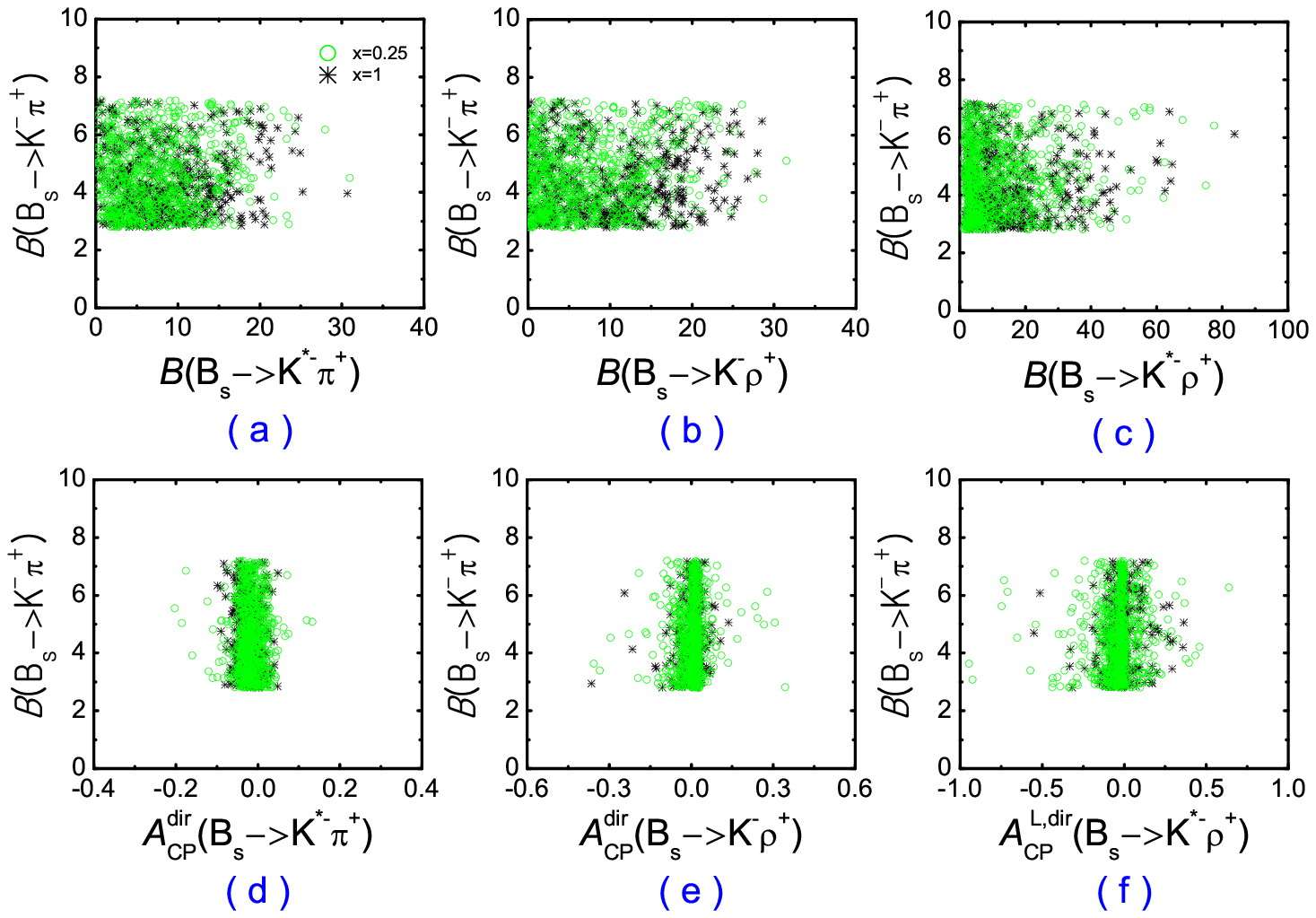}
\end{center}\vspace{-0.8cm}
 \caption{\small For the LR insertion, the
 correlation plots
between $\mathcal{B}(B_s\to K^{-}\pi^+)$ and $\mathcal{B}(B_s\to
K^{*-}\pi^+,K^{(*)-}\rho^{+})$ are shown in (a-c), and the
correlation plots between $\mathcal{B}(B_s\to K^{-}\pi^+)$ and
$\mathcal{A}^{dir}_{CP}(B_s\to K^{*-}\pi^+,K^{(*)-}\rho^{+})$ are
shown in (d-f). }
 \label{BBA}
\begin{center}
\includegraphics[scale=0.8]{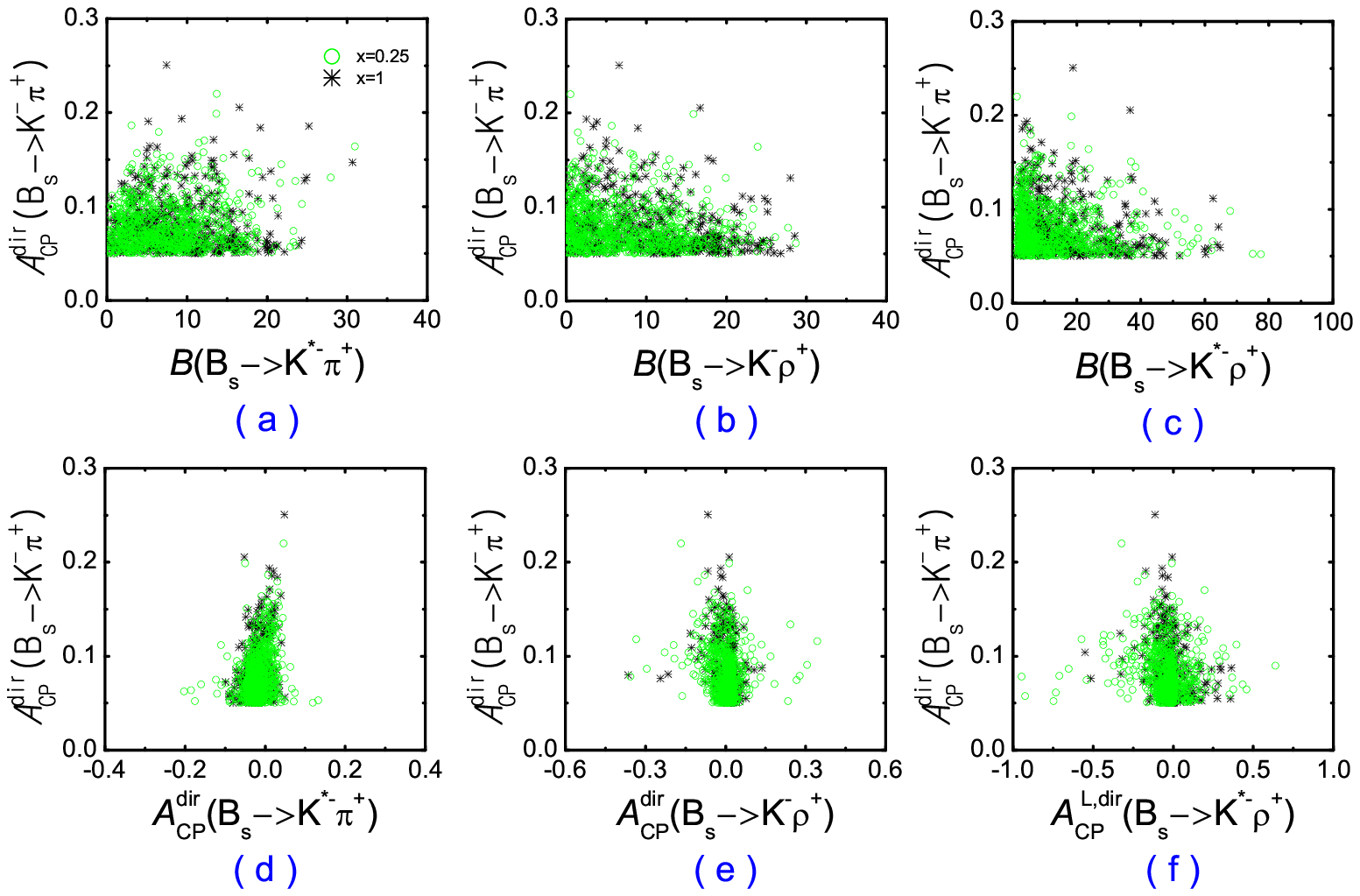}
\end{center}\vspace{-0.8cm}
 \caption{\small For the LR insertion, the
 correlation plots
between $\mathcal{A}^{dir}_{CP}(B_s\to K^{-}\pi^+)$ and
$\mathcal{B}(B_s\to K^{*-}\pi^+,K^{(*)-}\rho^{+})$ are shown in
(a-c), and the correlation plots between
$\mathcal{A}^{dir}_{CP}(B_s\to K^{-}\pi^+)$ and
$\mathcal{A}^{dir}_{CP}(B_s\to K^{*-}\pi^+,K^{(*)-}\rho^{+})$ are
shown in (d-f). }
 \label{ABA}
\end{figure}

In addition, for the LR MI case, we show the resulting predictions
for $\mathcal{B}(B_s\to K^{-}\pi^+)$ vs. $\mathcal{B}(B_s\to
K^{*-}\pi^+,K^{(*)-}\rho^{+})$ and $\mathcal{B}(B_s\to K^{-}\pi^+)$
vs. $\mathcal{A}^{dir}_{CP}(B_s\to K^{*-}\pi^+,K^{(*)-}\rho^{+})$ in
Fig. \ref{BBA}, as well as $\mathcal{A}^{dir}_{CP}(B_s\to
K^{-}\pi^+)$ vs. $\mathcal{B}(B_s\to K^{*-}\pi^+,K^{(*)-}\rho^{+})$
and $\mathcal{A}^{dir}_{CP}(B_s\to K^{-}\pi^+)$ vs.
$\mathcal{A}^{dir}_{CP}(B_s\to K^{*-}\pi^+,K^{(*)-}\rho^{+})$ in
Fig. \ref{ABA}. In all plots in Fig. \ref{BBA} and Fig. \ref{ABA},
the black and green points satisfy the constraints of $B_s\to
K^{-}\pi^+$ decay and  $B^0-\bar{B}^0$ mixing within $2\sigma$
ranges. As displayed in Fig. \ref{BBA}, both $\mathcal{B}(B_s\to
K^{*-}\pi^+,K^{(*)-}\rho^{+})$ and $\mathcal{A}^{dir}_{CP}(B_s\to
K^{*-}\pi^+,K^{(*)-}\rho^{+})$ are not very sensitive to the
constrained $\mathcal{B}(B_s\to K^{-}\pi^+)$. However, as shown in
Fig. \ref{ABA}, both $\mathcal{B}(B_s\to
K^{*-}\pi^+,K^{(*)-}\rho^{+})$ and $\mathcal{A}^{dir}_{CP}(B_s\to
K^{*-}\pi^+,K^{(*)-}\rho^{+})$ have some sensitivities to the
constrained $\mathcal{A}^{dir}_{CP}(B_s\to K^{-}\pi^+)$. Fig.
\ref{ABA} (a-c) show us, $\mathcal{A}^{dir}_{CP}(B_s\to K^{-}\pi^+)$
could has maximum when $\mathcal{B}(B_s\to
K^{*-}\pi^+,K^{(*)-}\rho^{+})$ is near the lower limits of the SM
predictions,  and  $\mathcal{A}^{dir}_{CP}(B_s\to K^{-}\pi^+)$ could
have smaller allowed range  with $\mathcal{B}(B_s\to
K^{*-}\pi^+,K^{(*)-}\rho^{+})$. Fig. \ref{ABA} (d-f) indicate that
there are some pionts accounting for large values of
$|\mathcal{A}^{dir}_{CP}(B_s\to K^{*-}\pi^+,K^{(*)-}\rho^{+})|$ when
$\mathcal{A}^{dir}_{CP}(B_s\to K^{-}\pi^+)$ is small.

There are similar correlations in the case of the RL MI as ones in
Fig. \ref{BBA} and Fig. \ref{ABA} expect the correlations between
$\mathcal{B}(B_s\to K^{-}\pi^+)$ and $\mathcal{A}^{dir}_{CP}(B_s\to
K^{-}\rho^+)$ as well as between $\mathcal{A}^{dir}_{CP}(B_s\to
K^{-}\pi^+)$ and $\mathcal{A}^{dir}_{CP}(B_s\to
K^{*-}\pi^+,K^{-}\rho^+)$.  $\mathcal{A}^{dir}_{CP}(B_s\to
K^{*-}\pi^+,K^{-}\rho^+)$ have no any sensitivities to
$\mathcal{A}^{dir}_{CP}(B_s\to K^{-}\pi^+)$  and $\mathcal{B}(B_s\to
K^{-}\pi^+)$.

\section{Conclusions}

Motivated by recent results from CDF, which favor a possible large
CP asymmetry and a small branching ratio in $B_s\rightarrow
K^-\pi^+$ decay, we have studied the gluino-mediated  SUSY
contributions with the MIs to four $B_s \to
K^{(*)-}\pi^{+},K^{(*)-}\rho^{+}$ decays based on the QCDF approach.
For the LL and RR MIs, we have found that the constrained  LL and RR
insertion effects from  $B^0-\bar{B}^0$ mixing are almost negligible
in $B_s \to K^{(*)-}\pi^{+},K^{(*)-}\rho^{+}$ decays, hence the LL
and RR MIs can not explain the measurements of $B_s\rightarrow
K^-\pi^+$ decay from CDF within $2\sigma$ ranges.
For the LR and RL MIs, we have fairly constrained  the LR and RL  MI
parameters from $B_s\to K^-\pi^+$ decay and  $B^0-\bar{B}^0$ mixing.
Furthermore, using the survived parameter spaces, we have explored
the LR and RL MI effects on the observables of three $B_s \to
K^{*-}\pi^{+},K^{(*)-}\rho^{+}$ decays, which have not been measured
yet.

The LR and RL insertions can generate sizable effects in $B_s \to
K^{(*)-}\pi^{+},K^{(*)-}\rho^{+}$ decays since their contributions
are enhanced by $m_{\tilde{g}}/m_b$. The allowed regions of both the
SUSY weak phases $\phi_{LR,RL}$ and the moduli
 $|(\delta_{LR,RL}^d)_{13}|$  have
been strongly constrained from $B_s\rightarrow K^-\pi^+$ decay and
$B^0-\bar{B}^0$ mixing within $2\sigma$ ranges.
We have found as long as $x\leq2$, there still is allowed spaces for
the LR and RL MI parameters. For $x=0.25$ case, $B_s\rightarrow
K^-\pi^+$ decay and $\Delta M_d$ provide the most stringent
constraint, however, for $x=1$ case, $B_s\rightarrow K^-\pi^+$ decay
and sin$2\beta$ provide the most stringent limit.
The theoretical predictions including the constrained LR and RL MI
contributions are compatible with the measurements within $2\sigma$
ranges from CDF collaboration in $B_s\rightarrow K^-\pi^+$ decay. We
have found the constrained LR and RL insertions still have obvious
effects on $\mathcal{B}(B_s\to K^{*-}\pi^+,K^{(*)-}\rho^{+})$,
$\mathcal{A}^{L,dir}_{CP}(B_s\to K^{*-}\rho^{+})$ and $f_L(B_s\to
K^{*-}\rho^{+})$, moreover, the LR insertion has obvious effects on
$\mathcal{A}^{dir}_{CP}(B_s\to K^{-}\rho^{+})$. Then we have
presented the sensitivities of the physical observable quantities to
the constrained LR and RL parameter spaces in Figs.
\ref{fig:MLR}-\ref{fig:PRL}. We have found
$\mathcal{A}^{dir}_{CP}(B_s\to K^-\rho^+, K^{*-}\rho^+)$ and
$f_L(B_s\to K^{*-}\rho^+)$ are very sensitive to the weak phases of
the  LR and RL insertion parameters. In addition, we also have shown
the
 correlations between two
observables of $B_s\rightarrow K^-\pi^+$ decay   and the observables
of $B_s\rightarrow K^{*-}\pi^+,K^{(*)-}\rho^{+}$ decays in Figs.
\ref{BBA}-\ref{ABA}. And we have found all observables of
$B_s\rightarrow K^{*-}\pi^+,K^{(*)-}\rho^{+}$ decays are not very
sensitive to $\mathcal{B}(B_s\to K^{-}\pi^+)$ for  the LR and RL
MIs, $\mathcal{B}(B_s\to K^{*-}\pi^+,K^{(*)-}\rho^{+})$ and
$\mathcal{A}^{dir}_{CP}(B_s\to K^{*-}\rho^{+})$ have some
sensitivities to $\mathcal{A}^{dir}_{CP}(B_s\to K^{-}\pi^+)$ in the
case of the LR and RL MIs, and  $\mathcal{A}^{dir}_{CP}(B_s\to
K^{*-}\pi^+,K^{-}\rho^{+})$ are also sensitive to
$\mathcal{A}^{dir}_{CP}(B_s\to K^{-}\pi^+)$ for the LR MI case.
The future measurements or precise measurements of the branching
ratios, the direct CP asymmetries and the polarization fractions in
$B_s \to K^{(*)-}\pi^{+},K^{(*)-}\rho^{+}$ decays could be used to
shrink/reveal/rule out the relevant LR and RL MI parameter spaces.
The results in this paper could be useful for probing SUSY effects
and searching direct SUSY signals at Tevatron and LHC in the near
future.

\end{document}